\documentclass[11pt,a4paper]{article}
\usepackage{jcappub}

\usepackage{graphicx}
\usepackage{mathrsfs}
\usepackage{url}
\usepackage{multirow}
\usepackage{array}
\usepackage{color}
\usepackage{amssymb,amsmath}

\begin{document}

\title{How CMB and large-scale structure constrain chameleon interacting dark energy}

\author[a]{Daniel Boriero,}
\author[b]{Subinoy Das,}
\author[c]{and Yvonne Y.~Y.~Wong}

\affiliation[a]{Fakult\"at f\"ur Physik, Universit\"at Bielefeld, Universit\"atstr. 25, Bielefeld, Germany}
\affiliation[b]{Indian Institute of Astrophisics, Bangalore, 560034, India}
\affiliation[c]{School of Physics, The University of New South Wales, Sydney NSW 2052, Australia}

\emailAdd{boriero@physik.uni-bielefeld.de, subinoy@iiap.res.in, yvonne.y.wong@unsw.edu.au}

\abstract{We explore a chameleon type of interacting dark matter--dark energy scenario in which a scalar field adiabatically traces the minimum of an effective potential sourced by the dark matter density. We discuss extensively the effect of this coupling on cosmological observables, especially the parameter degeneracies expected to arise between the model parameters and other cosmological parameters, and  then test the model against observations of the cosmic microwave background (CMB) anisotropies and other cosmological probes.  We find that the chameleon parameters $\alpha$ and $\beta$, which determine respectively the slope of the scalar field potential and the dark matter--dark energy coupling strength, can be constrained to $\alpha < 0.17$ and $\beta < 0.19$ using CMB data alone.  The latter parameter in particular is constrained only by the late Integrated Sachs--Wolfe effect.  Adding measurements of the local Hubble expansion rate $H_0$ tightens the bound on $\alpha$ by a factor of two, although this apparent improvement is arguably an artefact of the tension between the local measurement and the $H_0$ value inferred from Planck data in the minimal $\Lambda$CDM model.  The same argument also precludes chameleon models from mimicking a dark radiation component, despite a passing similarity between the two scenarios in that they both delay the epoch of matter--radiation equality.  
Based on the derived parameter constraints,  we discuss possible signatures of the model for  ongoing and future large-scale structure surveys.}
\maketitle


\section{Introduction}

There is a growing body of evidence that the universe is currently undergoing a phase of accelerated expansion (e.g.,~\cite{Riess:1998cb,Perlmutter:1998np,Suzuki:2011hu,Sherwin:2011gv,Ade:2013zuv,Planck:2015xua}). 
This apparent acceleration is usually attributed to a dark energy with an  equation of state $w \simeq -1$.  The fundamental nature of  dark energy, however, remains unknown.  At present, a cosmological constant or vacuum energy with $w=-1$ appears to be simplest solution that can account for all available data.  But dynamical scalar field models of dark energy~\cite{Wetterich:1987fm,Ratra:1987rm,Caldwell:1997ii} or $f(R)$ modified gravity models \cite{Lombriser:2010mp,Dossett:2014oia,Bel:2014awa} remain viable, albeit highly constrained, possibilities.

Yet another interesting possibility are the interacting quintessence models~\cite{Amendola:1999er}.  Here, a scalar field responsible for driving the accelerated expansion interacts with dark matter and/or ordinary matter.%
\footnote{It has been shown that any $f(R)$ model can be mapped to an interacting quintessence model by a suitable conformal transformation of the metric to the Einstein frame. In  a sense, both $f(R)$ and interacting quintessence are scalar--tensor modifications of gravity that introduce a scalar fifth force~\cite{Faulkner:2006ub,Hu:2007pj,Bertschinger:2008zb}. }
In fact, in the absence of any symmetry forbidding the interaction, it is  quite natural to expect such couplings~\cite{Chimento:2003iea,Comelli:2003cv}.  An immediate consequence, however, is that the matter fields will perceive an additional long-range force on super-Mpc scales; in the case the scalar field couples to baryonic matter, any such ``fifth-force'' effect that may have trickled down to the sub-Mpc scales is  automatically subject to solar system constraints~\cite{Will:2001mx,Kapner:2006si} as well as bounds from laboratory test of gravity~\cite{Hoyle:2004cw,Yang:2012zzb}.

One possible way to avoid these local constraints  is the chameleon mechanism.  The basic idea is that because the scalar field interacts with matter, its mass is a function of the local matter density. In those chameleon dark energy models wherein the scalar field couples universally to all forms of matter, 
the high baryonic matter density on solar system and laboratory scales causes the scalar field to become heavy, thereby hiding the fifth force from local tests. This screening mechanism switches off on matter-thin super-Galactic scales, and ideally the fifth force should invert its behaviour so as to drive the accelerated expansion of the universe~\cite{Khoury:2003aq,Khoury:2003rn,Brax:2004qh,Brax:2005ew,Copeland:2006wr}.
However, a recent work~\cite{2012PhRvL.109x1301W} has shown that the range of the chameleon force in universally-coupled models is necessarily restricted to below the $\sim 1$~Mpc scale, so that on its own the chameleon field is in fact unable to account for the observed accelerated expansion.  This constraint can nonetheless be circumvented if the scalar field interacts only with non-baryonic matter, and the chameleon mechanism  as discussed above remains viable.

In this work, we consider one such interacting dark matter--dark energy (DM--DE) model proposed in~\cite{Das:2005yj},  in which the scalar field adiabatically tracks the minimum
of the effective potential starting from deep in the radiation-dominated era up to the present epoch.
This model is very simple in that the coupling between the dark matter spinor $\psi$ and the scalar field $\phi$ is of the Yukawa type, $f(\phi) \psi \bar{\psi}$, 
and the coupling function is chosen to be a positive exponential suppressed by the Planck scale, i.e., $f(\phi)= e^{\beta \phi/M_{\textrm{\tiny Pl}}}$, which is very common in the Einstein frame and emerges from many string theory models of compactification~\cite{Damour:1994zq};  the self-interaction potential for $\phi$ takes the runaway form $V(\phi) \sim \phi^{-\alpha}$.  Previously, a na\"{\i}ve estimate has put an upper limit on $\alpha$,  $\alpha \leq 0.2 $, while $\beta$, which controls the strength of the fifth force, was found to be practically unconstrained~\cite{Das:2005yj}.  In the present work,  we wish to confront this model with the most recent cosmological data, especially measurements of the cosmic microwave background (CMB) temperature anisotropies by the ESA Planck mission~\cite{Ade:2013zuv}, and reexamine the model's viability.

Our second motivation comes from recent hints of a possible excess of radiation energy density during the CMB decoupling epoch from measurements of the CMB damping tail by the Atacama Cosmology Telescope (ACT)~\cite{Dunkley:2010ge} and the South Pole Telescope (SPT)~\cite{Keisler:2011aw}, as well as 
from the combined analysis of Planck data and the Hubble parameter inferred in the local neighbourhood~\cite{Ade:2013zuv,Planck:2015xua}. 
Conventional explanations include a thermalised populatioin of eV-mass sterile neutrinos (e.g.,~\cite{Hamann:2010bk,Hamann:2011ge}) or other light particles (e.g.,~\cite{Vogel:2013raa}), 
relativistic decay products of heavy particles (e.g.,~\cite{Hasenkamp:2014hma,Bjaelde:2010vt,DiBari:2013dna}), as well as early dark energy~\cite{Linder:2008nq}.  However, ultimately, what the CMB anisotropies probe are the effects of the energy content on the evolution of the photon perturbations around the decoupling 
epoch~\cite{Hou:2011ec,Archidiacono:2013cha}.  To this end, chameleon models, in which the dark matter--dark energy interaction endows the dark matter component with a nonstandard time evolution that tends to delay the epoch of matter--radiation equality, may very well mimic the phenomenology of a radiation excess.

The plan of the paper is as follows. We begin in section~\ref{sec:review} with a brief review of the chameleon model, and present the relevant equations of motion.  In section~\ref{sec:lss} we discuss the effects of the chameleon model on cosmological observables, and the parameter degeneracies expected to arise between the model parameters $\alpha$ and $\beta$ and other cosmological parameters.  We perform a fit of the model to cosmological data in section~\ref{sec:results}, and discuss the implications for the chameleon model parameters as well as the possibility of chameleon dark energy mimicking dark radiation.  Using the results from section~\ref{sec:results}, we identify in section~\ref{sec:prev} possible observational signatures for future cosmological probes.  We state our conclusions in section~\ref{sec:conclusions}.


\section{Chameleon model} \label{sec:review}

We briefly review the chameleon model of~\cite{Das:2005yj} in this section, and present the relevant equations that determine the background evolution, the initial conditions, and as well as the evolution of the dark energy and dark matter perturbations. These equations will later be embedded into the Boltzmann code {\sc Camb}~\cite{Lewis:1999bs}, in order to analyse the impact of the chameleon mechanism on cosmological observables such as the CMB anisotropies and the large-scale structure distribution. 


\subsection{Background evolution} \label{sec:background}

Following~\cite{Brax:2004qh,Bean:2007ny}, we take the Einstein frame as the physical frame in which dark energy interacts with dark matter fields through a $\phi$-dependent conformally rescaled Jordan frame metric~$g_{\mu\nu}^{(i)}$, whose generic form is given by \begin{equation}
 g_{\mu\nu}^{(i)} = e^{2 \beta_i \phi/M_{\textrm{\tiny Pl}}} g_{\mu\nu} \ ,
\end{equation}
where $\beta_i$ are dimensionless coupling constants, and $M_{\textrm{\tiny Pl}}$ is the reduced Planck mass.
The matter spinor fields $\psi^{(i)}$ are understood to follow the geodesics of $g_{\mu\nu}^{(i)}$, while the original $g_{\mu\nu}=
{\rm diag} \left(-1, a^2, a^2,a^2 \right)$ is the flat Friedmann--Lema\^{\i}tre--Robertson--Walker (FLRW) metric in the Einstein frame with scale factor~$a$.
The action takes the form 
\begin{equation}
 S=\int d^4x \left\lbrace  \sqrt{-g} \left[ \frac{M_\textrm{\tiny Pl}^2}{2} \mathcal{R} -\frac{(\partial\phi)^2}{2} - V(\phi) \right] + \mathcal{L}_m(\psi^{(i)},g_{\mu\nu}^{(i)}) \right\rbrace \ ,
\end{equation}
where~$\mathcal{R}$ is the Ricci scalar, $\mathcal{L}_m$~the Lagrangian density for the matter fields, and  we have also added the kinetic and potential terms for the scalar field~$\phi$. Varying the action with respect to~$\phi$, we obtain an equation of motion for the scalar field: 
\begin{equation}
\ddot{\phi} + 3 H \dot{\phi}= -V_{,\phi}(\phi)+\sum_i \frac{\beta_i}{M_{\textrm{\tiny Pl}}}e^{4\beta_i\phi/M_{\textrm{\tiny Pl}}}g_{(i)}^{\mu\nu}T^{(i)}_{\mu\nu} \ , \label{eq:kleingordon}
\end{equation}
where an overhead ``$\cdot$'' denotes a derivative with respect to the cosmic time~$t$,  $H \equiv \dot{a}/a$ is the Hubble expansion rate,
$T^{(i)}_{\mu\nu} \equiv (-2/\sqrt{-g^{(i)}}) \delta \mathcal{L}_m/\delta g^{\mu \nu}_{(i)}$ the Jordan frame energy-momentum tensor of the~\textit{i}th matter field, and we have assumed that the scalar field is homogeneous and isotropic. 
Limiting the coupling to the dark sector,  equation~(\ref{eq:kleingordon}) reduces to
\begin{equation}
\label{eq:phieqn}
 \ddot{\phi}+3H\dot{\phi}=-V_{,\phi}(\phi)-\frac{\beta}{M_{\textrm{\tiny Pl}}} e^{4 \beta \phi/M_{\textrm{\tiny Pl}}} \tilde{\rho}_{\textrm{\tiny \tiny DM}} \ ,
 \end{equation}
where  $\tilde{\rho}_{\textrm{\tiny DM}} \equiv - g^{(\textrm{\tiny DM})}_{\mu \nu} T^{\mu \nu}_{(\textrm{\tiny DM})}$ is the Jordan frame mean dark matter energy density. 

To determine the evolution of  $\tilde{\rho}_{\textrm{\tiny DM}}$,  conservation of energy--momentum $T^{\mu (\textrm{\tiny DM})}_{\nu;\mu}= 0$ in the Jordan frame gives
\begin{equation}
 \dot{\tilde{\rho}}_{\textrm{\tiny DM}}+ 3 \left(H+ \frac{\beta}{M_{\textrm{\tiny Pl}}}  \dot{\phi} \right)\tilde{\rho}_{\textrm{\tiny DM}}=0\,
\end{equation}
which has the solution
\begin{equation}
\label{eq:dmeqn}
 \tilde{\rho}_{\textrm{\tiny DM}} = \frac{\tilde{\rho}_{\textrm{\tiny DM}}^{(0)}}{a^3}e^{-3\beta(\phi-\phi_0)/M_{\textrm{\tiny Pl}}} 
 = \frac{{\rho}_{\textrm{\tiny DM}}^{(0)}}{a^3}e^{-\beta (3 \phi+ \phi_0)/M_{\textrm{\tiny Pl}}}  \ ,
\end{equation}
where $\phi_0 \equiv \phi(a=1)$, and $\rho_{\textrm{\tiny DM}}^{(0)} \equiv \tilde{\rho}_{\textrm{\tiny DM}}^{(0)} e^{4 \beta \phi_0/M_{\textrm{\tiny Pl}}}$  
is identified as the present-day physical dark matter density in the Einstein frame.
Substituting equation~(\ref{eq:dmeqn}) into~(\ref{eq:phieqn}), we see immediately that 
the scalar field dynamics in this model is controlled by an effective potential 
\begin{equation}
 V_{\textrm{\tiny eff}} (\phi)= V(\phi)+ \frac{\rho_{\textrm{\tiny DM}}^{(0)}}{a^3}e^{\beta(\phi-\phi_0)/M_{\textrm{\tiny Pl}}} \ , \label{eq:veff}
\end{equation}
where, for this work, we adopt a simple power law runaway potential, 
\begin{equation}
\label{eq:pot}
 V(\phi) = M_\phi^4 \left( \frac{M_{\textrm{\tiny Pl}}}{\phi} \right)^\alpha \ ,
\end{equation}
with $\alpha$ a positive constant that characterises the scalar field's self-interaction, and~$M_\phi$ the scalar field mass. This form of potential is 
 common in many string compactification models~\cite{Damour:1994zq}, and has the property that the attractor solution possesses a growing DM--DE coupling.
In the limit $\alpha=0$, the model reduces to the case of a cosmological constant.

The Friedmann equation for the full system, including other standard forms of energy densities, reads
\begin{equation}
\label{eq:friedmann}
 H^2(t)=\left(\frac{\dot{a}}{a}\right)^2 = \frac{1}{3 M_{\textrm{\tiny Pl}}^2}\left[\frac{\rho_{\gamma}^{(0)}}{a^4}+\frac{\rho_{\nu}^{(0)}}{a^4}+\frac{\rho_{\textrm{\tiny B}}^{(0)}}{a^3}+\frac{\rho_{\textrm{\tiny DM}}^{(0)}}{a^3}e^{\beta(\phi-\phi_0)/M_{\textrm{\tiny Pl}}} + \frac{\dot{\phi}^2}{2}+V(\phi) \right] \ ,
\end{equation}
where 
 $\rho_{\gamma}^{(0)}$, $\rho_{\nu}^{(0)}$ and $\rho_{\textrm{\tiny B}}^{(0)}$ denote the present-day energy densities of 
 photons, massless neutrinos and baryons, respectively. Again, for a vanishing coupling~$\beta$ and a constant potential~$V(\phi)$, equation~(\ref{eq:friedmann}) reduces to the standard $\Lambda$CDM equation.   Observe that the DM--DE interaction induces a nonstandard time evolution for the dark matter density, i.e.,
 \begin{equation}
 \label{eq:dmevo}
 \rho_\textrm{\tiny DM}   = \frac{{\rho}_{\textrm{\tiny DM}}^{(0)}}{a^3}e^{\beta (\phi- \phi_0)/M_{\textrm{\tiny Pl}}}\ .
 \end{equation}
 Furthermore, while the equation of state for the scalar field {\it alone} is the canonical one, namely,
\begin{equation}
 w_\phi \equiv  \frac{P_\phi}{\rho_\phi}= \frac{\dot{\phi}^2/2 - V(\phi)}{\dot{\phi}^2/2 + V(\phi)} \ ,
\end{equation}
where $\rho_\phi$ and $P_\phi$ are, respectively, the scalar field's energy density and pressure, 
taking into account the nonstandard dark matter density~(\ref{eq:dmevo}),  the {\it effective} 
dark energy density (as far as the background expansion rate~$H(t)$ is concerned) is in fact
\begin{equation}
\label{eq:effde}
\rho_\textrm{\tiny DE} =\rho_\phi + \frac{\rho_\textrm{\tiny DM}^{(0)}}{a^3}\left(e^{\beta(\phi-\phi_0)/M_{\textrm{\tiny Pl}}} -1\right) \equiv
 (1-x) \rho_\phi,
\end{equation}
which has an effective equation of state 
\begin{equation}
 w_{\textrm{\tiny eff}} = w_\phi \times \left( 1-x \right)^{-1} \ . \label{eq:weff}
\end{equation}
Where the coupling increases with time, such as in our case, $x \geqslant 0$ holds, thereby driving 
$w_{\textrm{\tiny eff}}$~to a value  below $-1$,  and hence leading the dark energy to exhibit an apparent phantom behaviour.


\subsection{Attractor solution} \label{sec:attractor}

The attractor solution for chameleon models and its stability have been studied previously in~\cite{Bean:2007ny,Corasaniti:2008kx,Brax:2004qh}.
It is set by the minimum of the effective potential, i.e., $V_{\textrm{\tiny eff}}^{,\phi}(\phi_{\textrm{\tiny min}})=0$, and adiabatic evolution simply stipulates that $\phi(t)$ traces the minimum at all times.
For the potential~(\ref{eq:pot}), this means
\begin{equation}
f[\phi(t)] = \left(\frac{\phi_0}{\phi(t)}\right)^{\alpha +1} - \frac{e^{\beta (\phi(t)-\phi_0)/M_\textrm{\tiny Pl}}}{a^{3}(t)} \ \rightarrow 0 \ , \label{eq:bisect}
\end{equation}
which can be solved using a simple bisection algorithm once the scalar field's present-day value~$\phi_0$ has been determined.
The latter can be achieved by demanding that the field's present-day potential energy dominates over its kinetic energy, so that $\rho_\phi^{(0)} \approx V(\phi_0)$.  Then from $V_{\textrm{\tiny eff}}^{,\phi}(\phi_{\textrm{\tiny min}})=0$ we find  
\begin{equation}
\label{eq:phi0}
\frac{\phi_0}{M_{\textrm{\tiny Pl}} }\ \approx \ \frac{\alpha}{\beta} \times  \frac{\rho_\phi^{(0)}}{\rho_{\textrm{\tiny DM}}^{(0)}}  = 
 \frac{\alpha}{\beta} \left( \frac{h^2}{\omega_\textrm{\tiny DM}} -1 \right)\ , 
\end{equation}
where the last equality in terms of the reduced density parameter~$\omega_\textrm{\tiny DM}$ and Hubble parameter~$h$
 follows from the assumption of a flat spatial geometry.
 An estimate for the mass scale of $\phi$~field,
\begin{equation}
\label{eq:fieldmass}
 M_\phi  \approx \left[ \rho_\phi^{(0)}  \left(\frac{\phi_0}{M_{\textrm{\tiny Pl}}} \right)^\alpha  \right]^{1/4} \  ,
\end{equation}
also follows immediately.
Thus, once~$\omega_{\textrm{\tiny DM}}$ an~$h$
have been specified,
it is possible to uniquely parameterise an adiabatically evolving chameleon model with only two parameters~$\alpha$ and~$\beta$ in a spatially flat universe.

\begin{figure}
\begin{center}
{\resizebox{1\columnwidth}{!}{\includegraphics[trim={0 0 0 0cm},clip]{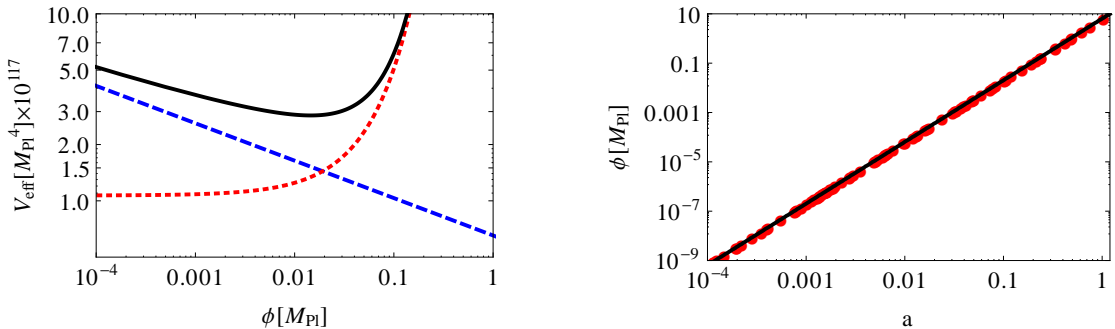}}}
 \end{center}
\vspace{-0.8cm}
\caption{{\it Left}: The effective potential $V_\textrm{\tiny eff}$ (solid black) for the model parameters $(\alpha, \beta,\omega_\textrm{\tiny DM},h) = {(0.2, 0.1, 0.13, 0.7)}$.  It is a composite of the original potential $V(\phi)$ (dashed blue) and the coupling with dark matter (dotted red). {\it Right}: The corresponding $\phi$~field as a function of the scale factor~$a$, obtained from numerically solving equation~(\ref{eq:phieqn}) (dotted red), and from the analytical solution~(\ref{eq:bisect}) (solid black).
}\label{figch1}
\end{figure}

Figure~\ref{figch1} shows the effective potential $V_\textrm{\tiny eff}(\phi)$ for choice of parameters $(\alpha, \beta,\omega_\textrm{\tiny DM},h) = {(0.2, 0.1, 0.13, 0.7)}$, and the corresponding evolution of $\phi$---the numerical solution of equation~(\ref{eq:phieqn}) as well as the analytical 
solution~(\ref{eq:bisect})---as a function of the scale factor~$a$.  Note that we have chosen the initial value of the scalar field $\phi^{\textrm{\tiny in}}$ to be in the 
undershooting regime, i.e., $0<\phi^{\textrm{\tiny in}}<\phi_{\textrm{\tiny min}}^{\textrm{\tiny in}}$.  
This ensures the satisfaction of the slow-roll conditions,
\begin{equation}
\label{eq:slowroll}
V_{,\phi\phi}^{1/2} \gg H ,\qquad
\dot{\phi}^2 \ll 2 V(\phi),
 \end{equation}
at all times, and consequently the stability of the linear density perturbations~\cite{Corasaniti:2008kx}.

An immediate phenomenological consequence of the attractor solution is that the effective dark energy equation of state~$w_{\textrm{\tiny eff}}$ as defined in equation~(\ref{eq:weff}), and consequently the cosmological background evolution, 
can have no dependence on the coupling parameter $\beta$.  This is easy to see by first defining $\tilde{\phi} \equiv \beta \phi/M_{\textrm{\tiny Pl}}$.  Then 
the present-day field value is given by $\tilde{\phi}_0 \approx \alpha (h^2/\omega_{\textrm{\tiny DM}}-1)$ according to equation~(\ref{eq:phi0}).  Substituting this into the attractor solution~(\ref{eq:bisect}), we see straight away that $\tilde{\phi}(t)$ does not depend on~$\beta$.  It follows then that $w_\textrm{\tiny eff}$, whose modification due to the DM--DE coupling comes in through the exponential  $e^{\tilde{\phi}-\tilde{\phi}_0}$, is also independent of~$\beta$.   To measure $\beta$ we need to consider its effects on the evolution of the inhomogeneities.


\subsection{Evolution of inhomogeneities} \label{sec:perturbation}

The presence of a DM--DE coupling likewise modifies the evolution of the inhomogeneities in the universe.  Working in the synchronous gauge whose line element is
\begin{equation}
ds^2 = -dt^2 + a^2(t) (\delta_{ij}+\tilde{h}_{ij}) dx^i dx^j\, ,
\end{equation}
the equations of motion for the dark matter density contrast $\delta_\textrm{\tiny DM}(k,t)$ and velocity divergence~$\theta_\textrm{\tiny DM}(k,t)$ for Fourier mode~$k$ 
are (e.g.,~\cite{Corasaniti:2008kx})
\begin{equation} 
\begin{aligned}
\dot{\delta}_{\textrm{\tiny DM}}&=-\left( \frac{\theta_{\textrm{\tiny DM}}}{a} + 
\frac{\dot{\tilde{h}}}{2} \right) + \frac{\beta}{M_\textrm{\tiny Pl}} \delta \dot{\phi} \ , \\
\dot{\theta}_{\textrm{\tiny DM}} &= -H\theta_{\textrm{\tiny DM}}+\frac{\beta}{M_\textrm{\tiny Pl}}\left( \frac{k^2}{a}\delta\phi -\dot{\phi}\theta_{\textrm{\tiny DM}} \right) \ , \label{eq:coldvelocity}
\end{aligned}
\end{equation} 
and similarly for the scalar field perturbations,
\begin{equation} 
\ddot{\delta \phi} + 3H\dot{\delta \phi}+\left( \frac{k^2}{a^2} +V_{,\phi\phi} \right)\delta\phi + \frac{1}{2}\dot{\tilde{h}}\dot{\phi}=-\frac{\beta}{M_\textrm{\tiny Pl}} \rho_{\textrm{\tiny DM}}\delta_{\textrm{\tiny DM}} \ ,  \label{eq:scalarperturbations}
\end{equation}  
where $\tilde{h} \equiv \tilde{h}^i_{\ i}$ is the trace of metric perturbation~$\tilde{h}_{ij}$.
Note that in non-interacting models the dark matter velocity divergence~$\theta_\textrm{\tiny DM}$ is exactly vanishing at all times because the synchronous coordinates are defined such that initially
$\dot{\theta}^{\textrm{\tiny in}}_{\textrm{\tiny DM}} =\theta^{\textrm{\tiny in}}_{\textrm{\tiny DM}} =0$, and there is no source term to perturb $\theta_\textrm{\tiny DM}$ away from zero.
In chameleon models, however, the DM--DE coupling engenders a source term proportional to the scalar field perturbation~$\delta \phi$, so that it becomes necessary to 
track the evolution of $\theta_\textrm{\tiny DM}$ as well.

The equation of motion for the metric perturbation~$\tilde{h}$, obtained from the perturbed Einstein equation,  likewise needs to be modified according to
\begin{eqnarray}
\label{eq:metricperturb}
 k^2 \eta -\frac{1}{2} a^2 H \dot{\tilde{h}} &=& -\frac{a^2}{2 M^2_\textrm{\tiny Pl}} \left(\sum_{i=\gamma,\nu,\textrm{\tiny B},\textrm{\tiny DM}}
\rho_i \delta_i +  \delta \rho_\phi \right), \\
 k^2 \dot{\eta} &=& \frac{a}{2 M^2_\textrm{\tiny Pl}} \left( \sum_{i=\gamma,\nu,\textrm{\tiny B},\textrm{\tiny DM}} (\rho_i+ P_i) \theta_i + 
a k^2 \dot{\phi} \delta \phi \right) \ , \\
\ddot{\tilde{h}} + 3 H  \dot{\tilde{h}} - 2 \frac{k^2}{a^2} \eta & =&  - \frac{3}{M_\textrm{\tiny Pl}^2} \left(\sum_{i=\gamma,\nu,\textrm{\tiny B},\textrm{\tiny DM}}
 \delta P_i  + \dot{\phi}\delta\dot{\phi}-V_{,\phi}\delta\phi  \right), \label{eq:metricperturb2}
\end{eqnarray}
where $\delta \rho_\phi = \dot{\phi}\delta\dot{\phi}+V_{,\phi}\delta\phi$, and 
we observe that perturbations in the $\phi$ field do not contribute to the traceless part of the spatial stress tensor.  
Note that the evolution equations presented in this section are common to {\it all}
scalar-field-based couple DM--DE models;  the uniqueness of the chameleon model lies in the behaviour of the potential and the existence of an attractor solution.

It is instructive to rewrite equation~(\ref{eq:coldvelocity}) as a second order differential equation for~$\delta_\textrm{\tiny DM}$,
\begin{equation}
\label{eq:delta2}
\ddot{\delta}_\textrm{\tiny DM} + 2 H \dot{\delta}_\textrm{\tiny DM} + \frac{1}{2} \left( \ddot{\tilde{h}} + 2 H \dot {\tilde{h}} \right) \approx \frac{\beta^2}{M_\textrm{\tiny Pl}^2} \frac{\rho_\textrm{\tiny DM} \delta_\textrm{\tiny DM}}{1+a^2 V_{,\phi \phi}/k^2}\ ,
\end{equation}
where we have ignored terms proportional to $\dot{\phi}$ because of the slow-roll conditions~(\ref{eq:slowroll}), and solved  equation~(\ref{eq:scalarperturbations}) for $\delta \phi$
assuming a steady state (i.e., $\ddot{\delta \phi} = \dot{\delta \phi} =0$).  
While the term dependent on the metric perturbation~$\tilde{h}$ arises from standard Einsteinian gravity
and can be easily constructed from equations~(\ref{eq:metricperturb}) and~(\ref{eq:metricperturb2}), the additional term proportional to $\beta^2$ can be interpreted as a fifth force acting on the dark matter perturbations.  The (comoving) range of this fifth force is determined by 
\begin{equation}
\lambda_\textrm{\tiny F} (a) \equiv a^{-1} V_{,\phi \phi}^{-1/2}\ ,
\end{equation}
so that only those scales satisfying $k > \lambda_\textrm{\tiny F}^{-1}$ will feel the force's effect.
For potentials of the form~(\ref{eq:pot}), we find
\begin{equation}
\label{eq:lambdaF}
\lambda_\textrm{\tiny F} (a) = a^{-1} \sqrt{\frac{\phi^{\alpha+2}}{\alpha (\alpha+1) M_\phi^4 M_\textrm{\tiny Pl}^{\alpha}}} 
 \approx a^2 H_0^{-1} \sqrt{ \frac{\Omega_\phi}{\Omega_\textrm{\tiny DM}^2} \frac{\alpha}{3\beta^2}} e^{-\beta(\phi-\phi_0)/M_\textrm{\tiny Pl}}\ ,
\end{equation}
where we have assumed in the last equality~$\alpha \ll 1$.  Thus, while $\lambda_\textrm{\tiny F}$ today is comparable to the present-day Hubble length, at early times the range of the force is strongly suppressed by the scale factor~$a$.  At the time of CMB decoupling, for example, equation~(\ref{eq:lambdaF}) evaluates to
\begin{equation}
\lambda_\textrm{\tiny F} (a_*) 
 \approx a^{3/2}_* (a_* H_*)^{-1} \sqrt{ \frac{\Omega_\phi}{\Omega_\textrm{\tiny DM}} \frac{\alpha}{3\beta^2}} e^{-\beta(\phi-\phi_0)/M_\textrm{\tiny Pl}}\ ,
\end{equation}
which shows that $\lambda_\textrm{\tiny F}(a_* \sim 0.001)$ is no more than about $10^{-5}$ times the comoving Hubble length at~$a_*$, and consequently completely out of the observable range of the CMB primary anisotropies ($> \mathcal{O}( 0.1) \times (a_* H_*)^{-1}$).


\section{Effects on cosmological observables} \label{sec:lss}

We describe in this section the observational consequences of the chameleon model for the current generation of cosmological probes.  In particular, we discuss the parameter degeneracies expected to arise between the $\alpha$ and $\beta$ and other cosmological parameters, and determine which combinations of cosmological observations would be capable of lifting these degeneracies.  We defer  the discussion of chameleon effects for  future observations  to section~\ref{sec:prev}.

\subsection{CMB anisotropies}
\label{sec:cmb}

Because the DM--DE coupling causes the dark matter density to dilute more slowly than $a^{-3}$, for the same present-day $\rho_\textrm{\tiny DM}^{(0)}$ value, the chameleon model has a lower dark matter density at high redshifts than does the standard $\Lambda$CDM case.  This has the effect of delaying the epoch of matter--radiation equality as per
\begin{equation}
\begin{aligned}
\label{eq:zeq}
1+z_\textrm{\tiny eq} &\approx \frac{\omega_\textrm{\tiny DM} e^{-\beta \phi_0/M_\textrm{\tiny Pl}} + \omega_\textrm{\tiny B}}{\omega_\gamma} \frac{1}{1 + 0.227 N_\textrm{\tiny eff}} 
\\
&\approx \frac{\omega_\textrm{\tiny DM} e^{-\alpha (h^2/\omega_\textrm{\tiny DM}-1 )} + \omega_\textrm{\tiny B}}{\omega_\gamma} \frac{1}{1 + 0.227 N_\textrm{\tiny eff}} 
\ ,
\end{aligned}
\end{equation}
where $N_\textrm{\tiny eff} = 3.046$, and the second approximate equality follows from equation~(\ref{eq:phi0}).
Changing the epoch of equality has a profound impact on the odd acoustic peak ratios of the CMB temperature anisotropies, and with seven  acoustic peaks now observed by Planck, $z_\textrm{\tiny eq}$ and hence $\omega_\textrm{\tiny DM} e^{-\alpha (h^2/\omega_\textrm{\tiny DM}-1 )}$ can be considered a well-constrained quantity.%
\footnote{
The baryon density~$\omega_\textrm{\tiny B}$ is independently well constrained by the odd-to-even acoustic peak ratios.
}

However, merely measuring $\omega_\textrm{\tiny DM} e^{-\alpha (h^2/\omega_\textrm{\tiny DM}-1 )}$ is clearly not sufficient to determine all of the chameleon model parameters, because of the intricate dependence of the exponent on the Hubble parameter $h$, the potential's slope~$\alpha$, as well as $\omega_\textrm{\tiny DM}$ itself, let alone the missing~$\beta$.
Breaking the degeneracy requires that we measure three other parameter combinations; two of these come automatically from the CMB temperature anisotropies:

\paragraph{Angular sound horizon.} The $\theta_\textrm{\tiny s} = r^*_\textrm{\tiny s}/D^*_\textrm{\tiny A}$ parameter determines the positions of the CMB acoustic peaks.  In the numerator, $r_\textrm{\tiny s}^*  \equiv r_\textrm{\tiny s}(a_*)= \int_0^{t_*} d t  \: c_\textrm{\tiny s}(t)/a(t)$ is the comoving sound horizon at the time of CMB decoupling.  For the chameleon model this evaluates, up to a constant factor, to
\begin{equation}
\label{eq:rs}
r_\textrm{\tiny s}^* \propto \sqrt{\frac{4}{3} \frac{a_\textrm{\tiny eq}}{(\omega_\textrm{\tiny DM} e^{-\alpha (h^2/\omega_\textrm{\tiny DM}-1 )} + \omega_\textrm{\tiny B})
 R(z_\textrm{\tiny eq})}} \ln \left[ \frac{\sqrt{1+R (z_*)} + \sqrt{R(z_*)+R(z_\textrm{\tiny eq})}}{1+\sqrt{R(z_\textrm{\tiny eq})}} \right] \ ,
\end{equation}
where $R (z) \equiv(3/4) (\omega_\textrm{\tiny B}/\omega_\gamma) a$.  In the denominator, $D^*_\textrm{\tiny A} \equiv D_\textrm{\tiny A}(a_*)$~is the angular diameter distance to the last scattering surface, 
which in the chameleon model takes the approximate form
\begin{equation}
\label{eq:da}
D^*_\textrm{\tiny A} \propto \int^1_{a_*} \frac{d a}{a^2 \sqrt{(\omega_\textrm{\tiny DM} +\omega_\textrm{\tiny B}) a^{-3} + (h^2 - \omega_\textrm{\tiny DM}-\omega_\textrm{\tiny B})e^{-3\int_1^a d\textrm{ln}a' (1+w_\textrm{\tiny eff})}}} \ ,
\end{equation}
up to a constant factor.
  Measuring~$z_\textrm{\tiny eq}$ and~$\omega_\textrm{\tiny B}$ from the peak ratios completely fixes~$r^*_\textrm{\tiny s}$, but not $D^*_\textrm{\tiny A}$ which depends on a different combination of~$\omega_\textrm{\tiny DM}$ and~$h$, as well as~$\alpha$  through the effective dark energy equation of state~$w_\textrm{\tiny eff}$
 (as already discussed in section~\ref{sec:attractor}, $w_\textrm{\tiny eff}$~does not depend on~$\beta$).   Therefore, $\theta_\textrm{\tiny s}$ contributes towards breaking the 
($\alpha, h, \omega_\textrm{\tiny DM})$-degeneracy.

\paragraph{Late Integrated Sachs--Wolfe (ISW) effect.}  The late ISW effect is primarily sensitive to the evolution of metric perturbations
on those length scales that enter the Hubble horizon while the universe transitions from matter to dark energy domination at low redshifts, $z\lesssim 1$.
The metric perturbations evolve according to equations~(\ref{eq:metricperturb}) to~(\ref{eq:metricperturb2})  in the chameleon model, and have contributions from both the matter component and the scalar field.

\begin{figure}[t]
\begin{center}
{\resizebox{1\columnwidth}{!}{\includegraphics{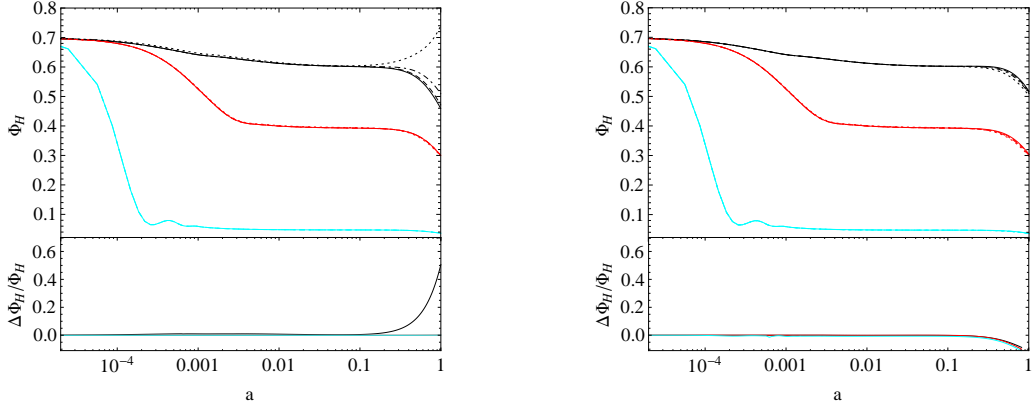}}}
\end{center}
\vspace{-0.6cm}
\caption{Bardeen potential $\Phi_H(k)$ as a function of the scale factor~$a$ for $k=0.0001 \  h\, {\rm Mpc}^{-1}$ (black), $0.01 \  h\, {\rm Mpc}^{-1}$ (red), and $0.1 \  h\, {\rm Mpc}^{-1}$ (cyan).  {\it Left}: Chameleon models with a fixed $\alpha = 0.1$, and $\beta$~values of 0.01 (solid), 0.05 (dashed), 0.1 (dot-dashed) and 0.2 (dotted); the bottom portion shows the fractional difference $\Delta \Phi_H/\Phi_H$ between the $\beta=0.01$ and $0.2$ cases.
{\it  Right}: Chameleon models with a fixed $\beta = 0.1$, and $\alpha$~values of 0.01 (solid), 0.05 (dashed), 0.1 (dot-dashed) and 0.2 (dotted); the bottom portion shows the fractional difference between the $\alpha=0.01$ and 0.2 cases.  In both panels
all other cosmological parameters have been fixed to the $\Lambda$CDM best-fit values from Planck+WP+HST.}\label{figbardeen}
\end{figure}

The scalar field  density perturbation $\delta \rho_\phi$ is given approximately by
\begin{equation}
\delta \rho_\phi \approx V_{,\phi} \delta \phi \approx 3  \beta^2 \Omega_\textrm{\tiny DM}(t)   \frac{a^2 H^2(t)}{k^2}\frac{\rho_\textrm{\tiny DM}} 
{1+a^2 V_{,\phi \phi}/k^2} 
\delta_\textrm{\tiny DM}\, , \label{eq:steadystate}
\end{equation}
where we have used $V_{,\phi} = -(\beta/M_\textrm{\tiny Pl}) \rho_\textrm{\tiny DM}$, and assumed the steady-state solution of equation~(\ref{eq:scalarperturbations}) for $\delta \phi$ (i.e., $\ddot{\delta \phi} = \dot{\delta \phi} = 0$).    Feeding this solution into equations~(\ref{eq:metricperturb}) and~(\ref{eq:metricperturb2}) also allows us to rewrite the evolution
equation~(\ref{eq:delta2}) for the matter perturbations in the low-redshift universe as 
\begin{equation}
\label{eq:deltadm2}
\ddot{\delta}_\textrm{\tiny DM} + 2 H \dot{\delta}_\textrm{\tiny DM} = \frac{3}{2} H^2 \Omega_\textrm{\tiny DM}(t) \left[ 1+
 \frac{ 2 \beta^2}{1+a^2 V_{,\phi \phi}/k^2} \left(1 - 3   \Omega_\textrm{\tiny DM}(t)   \frac{a^2 H^2(t)}{k^2} \right) \right] \delta_\textrm{\tiny DM}\ ,
\end{equation}
 where we have neglected the baryon component for simplicity.
Since according to equation~(\ref{eq:lambdaF}) the range of the fifth force has recently attained Hubble length, we can establish that
$a^2 V_{,\phi \phi}/k^2 \lesssim 1$ for all observable scales at $z \lesssim 1$.  It then follows from equation~(\ref{eq:steadystate}) 
that $\delta \rho_\phi$ is strongly dependent on the coupling parameter~$\beta$, and its impact on the metric perturbations---both directly and indirectly through 
$\delta_\textrm{\tiny DM}$---is greatest on scales close to the Hubble length due to its $a^2 H^2(t)/k^2$ dependence.
We illustrate this point in figure~\ref{figbardeen}  with the gauge-invariant Bardeen potential,
\begin{equation}
\Phi_H \equiv\frac{a^2}{2 M_\textrm{\tiny Pl}^2 k^2}  \left\{ \sum_{i=\gamma,\nu,\textrm{\tiny B},\textrm{\tiny DM}} \left[
\rho_i \delta_i + \frac{3 a H}{k^2} (\rho_i+P_i) \theta_i \right]
+  \delta \rho_\phi  + 3 a^2 H \dot{\phi} \delta \phi
\right\},
\end{equation}
as a function of the scale factor~$a$ at $k= 0.0001, 0.01, 0.1 \ h\,  {\rm Mpc}^{-1}$  for various combinations of $\alpha$ and~$\beta$.  

A second interesting feature in equations~(\ref{eq:steadystate}) and~(\ref{eq:deltadm2}) is that, while the low-redshift behaviour of~$H(t)$ can be conveniently reparameterised  in terms of an effective dark energy equation of state~$w_\textrm{\tiny eff}$, the time-dependent reduced density parameter $\Omega_\textrm{\tiny DM}(t)$, given in this case by
\begin{equation}
\label{eq:omegadmt}
\Omega_\textrm{\tiny DM}(t) \equiv  \frac{\rho^{(0)}_\textrm{\tiny DM} a^{-3}}{\rho^{(0)}_\textrm{\tiny DM} a^{-3} +\rho_\phi e^{-\beta(\phi-\phi_0)/M_\textrm{\tiny Pl}}} \ ,
\end{equation}
does not share the same mapping, i.e., $ \rho_\phi e^{-\beta(\phi-\phi_0)/M_\textrm{\tiny Pl}}$ in the demonimator does not equate to $\rho_\textrm{\tiny DE}$ of equation~(\ref{eq:effde}).  Thus, even if the $\beta^2$ term should be absent in equation~(\ref{eq:deltadm2}), the evolution of $\delta_\textrm{\tiny DM}$ and hence $\Phi_H$ in the chameleon model {\it cannot}  be  mapped to that in a non-interacting scenario with a dark energy equation of state~$w_\textrm{\tiny eff}$.  In other words, the evolution of $\delta_\textrm{\tiny DM}$ and $\Phi_H$ in the low-redshift universe probes a combination of $(\alpha, h, \omega_\textrm{\tiny DM})$ different from that determined by $w_\textrm{\tiny eff}$.

\bigskip

Note that while measurements of the CMB angular damping scale~$\theta_\textrm{\tiny d} = r^*_\textrm{ d}/D^*_\textrm{\tiny A}$ by ACT~\cite{Dunkley:2010ge}
and~SPT~\cite{Keisler:2011aw} have been instrumental in procuring the first CMB-only constraints on the radiation energy density, for the chameleon model $\theta_\textrm{\tiny d}$ offers no new information beyond what can already be obtained from the acoustic peaks.  To see this, we note that the diffusion damping scale~$r^*_\textrm{\tiny d} \equiv r_\textrm{\tiny d}(a_*)$ can be approximated as~\cite{Hou:2011ec}
\begin{equation}
\begin{aligned}
{r^*_\textrm{\tiny d}}^2 &\approx (2 \pi)^2 \int_0^{a_*} \frac{da}{a^3 \sigma_T n_e H} \left[ \frac{R^2 + (16/15) (1+R)}{6 (1+R^2)} \right] \\
& \propto  \frac{(2 \pi)^2}{\sqrt{\omega_\textrm{\tiny DM} e^{-\alpha (h^2/\omega_\textrm{\tiny DM}-1 )}
+ \omega_\textrm{\tiny B}}} \int_0^{a_*} \frac{da}{a \sqrt{a + a_{\rm eq}} \sigma_T n_e} \left[ \frac{R^2 + (16/15) (1+R)}{6 (1+R^2)} \right] \ ,
\end{aligned}
\end{equation}
where $\sigma_T$ is the Thomson scattering cross-section, and $n_e$ is the free electron number density.  
Clearly, after~$z_\textrm{\tiny eq}$ and~$\omega_\textrm{\tiny B}$ have been fixed, $r^*_\textrm{\tiny d}$~is likewise completely determined, while the angular diameter distance~$D^*_\textrm{\tiny A}$ is probed by the angular sound horizon scale~$\theta_\textrm{\tiny s}$.  The information contained in the angular diffusion scale~$\theta_\textrm{\tiny d}$ is therefore redundant.  Indeed, as we shall see in section~\ref{sec:results}, adding ACT and SPT data to WMAP's acoustic peak measurements does not improve the constraints on the chameleon model parameters.


\subsection{Non-CMB observables}

One more orthogonal parameter combination needs to be constrained in order to eliminate all degeneracies.  This can come from any one of the following observables:
\begin{enumerate}
\item The Hubble expansion parameter~$h$ can be fixed directly by measurements in our local neighbourhood.

\item The current generation of baryon acoustic oscillations (BAO)  measurements constrains the angular scale $r_\textrm{\tiny s}(z_\textrm{\tiny d})/D_\textrm{\tiny V}(z_\textrm{\tiny BAO})$, where $z_\textrm{\tiny d} \approx 1020$ denotes the redshift of the baryon drag epoch, $z_\textrm{\tiny BAO}$  the effective redshift at which the BAO  is observed, and 
\begin{equation}
D_\textrm{\tiny V} ( z) = \left[ z(1+z)^2 \frac{D_\textrm{\tiny A}^2(z)}{H(z)} \right]^{1/3}
\end{equation}
is a distance scale subsuming both the angular diameter distance $D_\textrm{\tiny A}(z)$ and the Hubble expansion rate
\begin{equation}
H(a) \propto 
 \sqrt{(\omega_\textrm{\tiny DM} +\omega_\textrm{\tiny B}) a^{-3} + (h^2 - \omega_\textrm{\tiny DM}-\omega_\textrm{\tiny B})e^{-3\int_1^a d\textrm{ln}a' (1+w_\textrm{\tiny eff})}} \ , .
\end{equation}
The BAO peak shares a common physical origin with the CMB acoustic peaks.  However, because the BAO measurement takes place at a low redshift ($z_\textrm{\tiny BAO} < 1$), the parameter combination to which it is sensitive is vastly different from that probed by the high-redshift $r_\textrm{\tiny s}^*/D_\textrm{\tiny A}^*$ of equations~(\ref{eq:rs}) and~(\ref{eq:da}).

\item An alternative low-redshift distance constraint comes from measurements of the luminosity distance $D_\textrm{\tiny L}(a)$ versus redshift using  type Ia supernovae. 
In a flat spatial geometry, $D_\textrm{\tiny L} (a) = D_\textrm{\tiny A}(a)/a^2$.

\item The large-scale matter power spectrum extracted from galaxy clustering surveys yields several useful pieces of information.  Firstly, the ``turning point'' depends on the comoving wavenumber
\begin{equation}
k_\textrm{\tiny eq} \equiv a_\textrm{\tiny eq} H(a_\textrm{\tiny eq}) \approx 4.7 \times 10^{-4} h^{-1} \sqrt{(\omega_\textrm{\tiny DM} e^{-\alpha (h^2/\omega_\textrm{\tiny DM}-1 )}
+ \omega_\textrm{\tiny B}) (1+z_\textrm{\tiny eq}})~ h \textrm{Mpc}^{-1} \ ,
\end{equation}
which, after fixing $z_\textrm{\tiny eq}$ and $\omega_\textrm{\tiny B}$ with the CMB, effectively depends only on the Hubble parameter~$h$ (recall that $k$ is measured in units of $h\textrm{Mpc}^{-1}$).  Secondly, BAO features are also present in the present generation of power spectrum  measurements (e.g.,~\cite{Reid:2009xm}), which offers the same handle on a low-redshift distance scale as discussed immediately above.  Thirdly, the fifth force induced by the DM--DE coupling in principle enhances clustering in a scale-dependent fashion even on scales well within the Hubble horizon according to  equation~(\ref{eq:deltadm2}), thereby distorting the matter power spectrum.  However, because the range of the force has only reached Hubble length at recent times, we expect this subhorizon spectral distortion to be confined mainly to the nonlinear scales ($k \gtrsim 0.2 \ h \textrm{Mpc}^{-1}$), where the density perturbations have had more time to evolve under the influence of the fifth force~\cite{Das:2005yj} and which are beyond the scope of our perturbative analysis. 
Nonetheless, we note that in certain $f(R)$ or interacting quintessence models it is possible to shift the clustering enhancement to the linear scales (e.g.,~\cite{Motohashi:2012wc,Brax:2011ta}).

\end{enumerate}

Other low-redshift matter distribution measurements such as the cluster mass function \cite{Reichardt:2012yj,Ade:2013skr,Mantz:2014paa}, cosmic shear \cite{Schrabback:2009ba,Jee:2012hr,Kitching:2014dtq}, and the Lyman-$\alpha$ forest \cite{McDonald:2004eu,Lee:2012xb} are also potentially useful for constraining chameleon cosmology
in that they provide a direct measurement of the density fluctuation amplitude which should be sensitive to the non-standard low-redshift evolution of the matter density perturbations induced by the DM--DE coupling.
However, to make use of these measurements requires nonlinear modelling, which is beyond the scope of this work.


\section{Cosmological data analysis} \label{sec:results}

Following the discussions of section~\ref{sec:lss}, we consider four generic types of cosmological measurements: temperature and polar polarisation power spectra of the CMB anisotropies, the BAO scale, the large-scale matter power spectrum, and direct measurements of the local Hubble expansion rate. 
The details of the actual data sets used are related below.  On these data sets we perform a Bayesian statistical inference analysis
using the publicly available Markov Chain Monte Carlo (MCMC) parameter estimation package {\sc CosmoMC}~\cite{Lewis:2002ah} coupled to the {\sc CAMB}~\cite{Lewis:1999bs} Boltzmann solver modified in accordance with section~\ref{sec:review}. With the exception of the local Hubble parameter measurement, the likelihood routines and the associated window functions are supplied by the experimental collaborations.

Table~\ref{tab1} shows the fit parameters used in the analysis and their associated (flat) priors ranges.  We adopt the Gelman--Rubin convergence criterion~$R-1< 0.02$ when generating our multiple Markov chains, where $R$ is the variance of chain means divided by the mean of chain variances.

\begin{table}[t]
\begin{center}
 {\footnotesize
 \vspace{2mm}
 \begin{tabular}{|l|p{9cm}|c|c|}\hline
Parameter 	 	& Description   			& Prior & $\Lambda$CDM \\ \hline 
$\omega_\textrm{\tiny B}$ 	& Baryon density   		& $0.005 \to 0.04$ & -- \\ 
$\omega_\textrm{\tiny DM} $ 	& Cold dark matter density   	& $0.01 \to 0.99$ & --  \\ 
$h$    	& Hubble parameter			& $0.4 \to 1.0$ & --\\ 
$z_\textrm{\tiny re}$   	& Redshift of reionization 		& $ 3 \to 35$ & -- 	 \\ 
$n_\textrm{\tiny s}$    	& Spectral index of the primordial power spectrum  & $0.5 \to 1.5$ & --\\ 
$\ln(10^{10} A_\textrm{\tiny s})$	& Amplitude of the primordial power spectrum at $k=0.05\ \textrm{Mpc}^{-1}$& $2.7 \to 4.0$ & --  \\ \hline
$\alpha$   	& Slope of the chameleon  potential & $10^{-5} \to 5$ & 0		\\ 
$\beta$   	& Strength of the chameleon coupling 	& $10^{-5} \to 5$ & 0 	\\  \hline
$N_{\textrm{\tiny eff}}$& Number of effective massless neutrino families & $1 \to 9.8$ & $3.046$  \\ \hline
 \end{tabular}
 }
 \end{center}
 \caption{Fitting parameters in our MCMC analysis and their associated flat prior ranges.  The six parameters in the first block are the standard $\Lambda$CDM parameters.  The chameleon model parameter space subsumes the $\Lambda$CDM parameter space, and has an additional two parameters~$\alpha$ and~$\beta$ which, when set to zero, 
reduce the chameleon model to $\Lambda$CDM.  The last entry~$N_\textrm{\tiny eff}$ parameterises the radiation excess; in $\Lambda$CDM this is fixed at $3.046$.\label{tab1}}
\end{table}


\subsection{Data sets}

We split our analysis into two stages, the first centred on the WMAP 9-year measurements of the cosmic microwave background temperature and polarisation anisotropies~\cite{Hinshaw:2012aka}, and the second on 
the more recent temperature measurements from the first data release of the Planck mission~\cite{Ade:2013zuv}.  We describe these and other auxiliary data sets below.

\paragraph{CMB anisotropies.}

In the stage 1, we use the temperature (TT), $E$-polarisation (EE), and $B$-polarisation (BB) autocorrelation power spectra, as well as the temperature--$E$-polarisation (TE) cross-correlation power spectra from WMAP~\cite{Ade:2013zuv}.  To this we add the TT power spectra from ACT and SPT~\cite{Dunkley:2013vu,Das:2013zf,Story:2012wx,Hou:2012xq,Calabrese:2013jyk}, which cover the range $600 < \ell < 3000$ (SPT) and $500 < \ell< 10000$ (ACT).  The ACT/SPT data include in addition the deflection power spectra ($\phi\phi$) due to gravitational lensing, which we also analyse in our study. 
In  stage 2, following the guidelines of~\cite{Ade:2013zuv}, we use the TT spectra from the Planck mission in conjunction with the WMAP 9-year polarisation measurements (WP).  To this we add the ACT/SPT TT spectrum at high multipoles (HighL), which helps to constrain the foreground nuisance parameters used in the Planck data analysis.

\paragraph{Local Hubble parameter measurements.}  
We adopt the values of $H_0$ measured in our local neighbourhood by the Hubble Space Telescope.  In stage 1, we use $H_0=74.2 \pm 3.6$ km s$^{-1}$ Mpc$^{-1}$~\cite{Riess:2009pu}, 
published in 2009, in conjunction with our analysis of the WMAP 9-year CMB measurements.  This value of $H_0$ was also adopted by the WMAP collaboration in their 9-year analysis~\cite{Hinshaw:2012aka}, and we opt to keep it to facilitate comparison.  In stage 2, we adopt a more recent measured value, $H_0=73.8 \pm 2.4$ km s$^{-1}$ Mpc$^{-1}$~\cite{2011ApJ...730..119R}, in combination with our Planck analysis. 
Because each value is used exclusively with either WMAP or Planck, we adopt the same acronym HST for both measurements.

\paragraph{Large-scale matter power spectrum}

We use the red luminous galaxy survey from the Sloan Digital Sky Survey DR-7 (SDSS)~\cite{Reid:2009xm} to constrain the matter power spectrum in the $k$-range $[0.02,0.2] \;  h {\rm Mpc}^{-1}$, in conjunction with the WMAP 9-year data in stage 1 of the analysis.
 The power spectrum measurement contains both broadband information on the scale-dependence, as well as geometric information in the form wiggles from the baryonic acoustic oscillations.  In combination with the WMAP CMB measurements, the latter information generally suffices to constrain model parameters that affect the late-time expansion of the universe, while the former is primarily useful for extended models in which the perturbations pick up an additional scale-dependence at late times (e.g., models with finite neutrino masses)~\cite{2010JCAP...07..022H}.  In the context of chameleon models, we expect geometric information to be the more useful of the two (the additional scale-dependence in chameleon models are in any case all confined to the nonlinear scales beyond $k \sim 0.2 \; h{\rm Mpc}^{-1}$, as discussed in section~\ref{sec:lss}).
 We therefore do not use the matter power spectrum in stage 2 of the analysis in combination with Planck data, and opt instead for direct measurements of the BAO scale.  See next.

\paragraph{BAO scale.}  We use measurements of the BAO scale by SDSS-DR~\cite{Padmanabhan:2012hf}, SDSS-DR9~\cite{Anderson:2012sa}, and 6dFGS~\cite{Beutler:2011hx}.   For cosmological models with no non-trivial scale dependence,  geometric information extracted from large-scale structure surveys suffices to constrain the model parameters~\cite{2010JCAP...07..022H}, and has the added advantage of being much less prone to the nonlinearity issues that plague broadband measurements of the large-scale matter power spectrum. 

\paragraph{Type Ia supernovae.} 

We include the supernovae data set Union2 compilation~\cite{2010ApJ...716..712A}, labelled ``SNIa'', in stage 1 of the analysis. The standardised luminosity of Type Ia supernovae is a classic measure of the redshift-dependence of the ($h$-normalised) luminosity distance. This compilation consists of 557 supernovae collected from different surveys, and was a standard compilation used also by the WMAP collaboration in their 9-year data analysis~\cite{Hinshaw:2012aka}.


\subsection{Parameter constraints}\label{sec:chamfits}

\begin{figure}[t]
\begin{center}
{\resizebox{0.95\columnwidth}{!}{\includegraphics{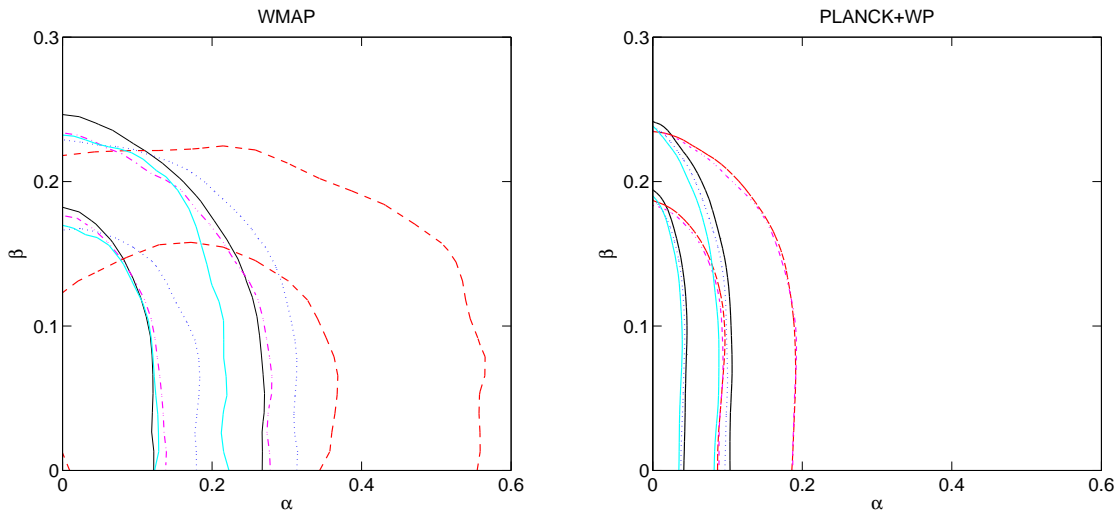}}}
\end{center}
\vspace{-0.6cm}
\caption{2D marginal 68\% and 95\% contours in the $(\alpha,\beta)$-plane derived from various data combinations. 
{\it Left}: WMAP temperature and polarisation measurements combined with HST (solid black), SDSS (dashed red), HST+SDSS (dotted blue), HST+ACT/SPT (dot-dashed magenta), and SDSS+SNIa (solid cyan). 
{\it Right}: Planck temperature and WMAP polarisation data combined with HST (solid black), BAO (dashed red), HighL+HST (dotted blue), HighL+BAO (dot-dashed magenta), and HighL+HST+BAO (solid cyan).}\label{ab1}
\end{figure}

Table~\ref{tab2} summarises the constraints on the chameleon model parameters~$\alpha$ and~$\beta$  from our MCMC analysis using various data combinations.  The corresponding two-dimensional marginal posteriors in the $(\alpha,\beta)$-plane are displayed in figure~\ref{ab1}.

\setlength{\tabcolsep}{3.3pt}
\renewcommand{\arraystretch}{1.1}

\begin{table}[t]
\begin{center}
{\footnotesize
 \begin{tabular}{|c|l|c|c|c|c|c|c|c|}\hline
 Model			& Data 	& $\alpha$ 		& $\beta$ 		& $N_{\textrm{\tiny eff}}$  & $\omega_\textrm{\tiny DM}$ & $h$ & $\Delta \chi_{\textrm{\tiny a}}^2/\Delta_\textrm{\tiny dof}$& $\Delta \chi_{\textrm{\tiny b }}^2/\Delta_\text{\tiny dof}$ \\ \hline
\parbox[t]{2mm}{\multirow{10}{*}{\rotatebox[origin=c]{90}{Chameleon}}}			
		& W+HST  							& $<0.23$ & $<0.19$ & $3.046$ & $0.109_{0.099}^{0.118}$ & $0.70_{0.65}^{0.75}$ & $+0.0/2$ & $+0.1/1$\\ \cline{2-9}
  		& W+HST+ACT/SPT 				& $<0.24$ & $<0.19$ & $3.046$ & $0.109_{0.100}^{0.119}$ & $0.70_{0.65}^{0.75}$ & $+1.3/2$ & $+0.7/1$ \\ \cline{2-9}
 			& W+SDSS 							& $<0.48$ & $<0.19$ & $3.046$ & $0.112_{0.103}^{0.121}$ & $0.65_{0.60}^{0.70}$ & $-0.1/2$ & $-0.3/1$ \\ \cline{2-9}
			& W+HST+SDSS 					& $<0.27$ & $<0.19$ & $3.046$ & $0.112_{0.104}^{0.120}$ & $0.68_{0.64}^{0.71}$ & $+0.1/2$ & $+1.1/1$ \\ \cline{2-9}
 			& W+SNIa+SDSS					& $<0.19$ & $<0.19$ & $3.046$ & $0.113_{0.105}^{0.120}$ & $0.68_{0.65}^{0.71}$ & $-0.4/2$ & $-0.3/1$ \\ \cline{2-9}
 			& P+WP+BAO						& $<0.17$ & $<0.19$ & $3.046$ & $0.118_{0.115}^{0.122}$ & $0.67_{0.64}^{0.69}$ & $+0.7/2$ & $+0.5/1$ \\ \cline{2-9}
 			& P+WP+HST						& $<0.09$ & $<0.19$ & $3.046$ & $0.117_{0.113}^{0.122}$ & $0.68_{0.65}^{0.71}$ & $+0.4/2$ & $+4.7/1$ \\ \cline{2-9}
 			& P+WP+HighL+BAO			& $<0.17$ & $<0.19$ & $3.046$ & $0.118_{0.115}^{0.122}$ & $0.67_{0.64}^{0.69}$  & $-0.7/2$ & $+0.4/1$ \\ \cline{2-9}
			& P+WP+HighL+HST			& $<0.09$ & $<0.19$ & $3.046$ & $0.117_{0.113}^{0.122}$ & $0.67_{0.64}^{0.69}$ & $-0.6/2$ & $+4.7/1$\\ \cline{2-9}
			& P+WP+HighL+BAO+HST	& $<0.08$ & $<0.19$ & $3.046$ & $0.117_{0.114}^{0.121}$ & $0.67_{0.65}^{0.69}$ & $-0.6/2$ & $+3.1/1$ \\ \hline
\parbox[t]{2mm}{\multirow{10}{*}{\rotatebox[origin=c]{90}{$N_{\textrm{\tiny eff}}\Lambda$CDM }}}
			& W+HST 		    				& -       & -  		  & $4.3_{2.8}^{6.1}$ & $0.13_{0.10}^{0.18}$	   & $0.75_{0.69}^{0.82}$ & $-0.1/1$ & -- \\ \cline{2-9}
      & W+HST+ACT/SPT				& -       & -       & $3.2_{2.6}^{3.9}$ & $0.14_{0.11}^{0.17}$ & $0.74_{0.69}^{0.79}$ & $+0.6/1$ & --\\ \cline{2-9}
      & W+SDSS 							& -	  	  & -	  		& $4.5_{2.1}^{7.6}$ & $0.14_{0.10}^{0.21}$   & $0.75_{0.62}^{0.91}$ & $+0.2/1$ & -- \\ \cline{2-9}
      & W+HST+SDSS 					& -   		& -  		  & $4.3_{3.1}^{5.5}$ 	& $0.14_{0.11}^{0.16}$	 & $0.74_{0.68}^{0.80}$   & $-1.0/1$ & -- \\ \cline{2-9} 
      & W+SNIa+SDSS 					& -   		& -  		  & $4.6_{2.4}^{7.4}$ & $0.14_{0.10}^{0.20}$	   & $0.76_{0.64}^{0.92}$  & $-0.1/1$ & --\\ \cline{2-9}
      & P+WP+BAO   				& -   		& -  		  & $3.3_{2.8}^{3.9}$ & $0.124_{0.114}^{0.134}$	   & $0.69_{0.66}^{0.73}$     & $+0.2/1$  & --\\ \cline{2-9}
      & P+WP+HST   				& -   		& -  		  & $3.7_{3.2}^{4.2}$ & $0.128_{0.118}^{0.137}$	   & $0.73_{0.69}^{0.77}$    & $-4.3/1$  & --\\ \cline{2-9}
      & P+WP+HighL+BAO 		& -   		& -   	  & $3.3_{2.7}^{3.8}$ & $0.123_{0.114}^{0.132}$	   & $0.65_{0.62}^{0.69}$   & $-1.1/1$ & -- \\ \cline{2-9}
      & P+WP+HighL+HST 		& -   		& -   		& $3.2_{2.7}^{3.6}$ & $0.119_{0.111}^{0.128}$   & $0.68_{0.65}^{0.72}$    & $-5.3/1$ & -- \\ \cline{2-9}
      & P+WP+HighL+BAO+HST & -   		& -   		& $3.2_{2.7}^{3.6}$ 	& $0.120_{0.112}^{0.128}$   & $0.68_{0.66}^{0.71}$    & $-3.8/1$ & -- \\ \hline
\end{tabular}
}
\end{center}
\caption{1D marginal statistics for the chameleon parameters~$\alpha$ and~$\beta$ in the chameleon fit (top block) and for~$N_\textrm{\tiny eff}$ in the $N_\textrm{\tiny eff}\Lambda$CDM fit (bottom block) derived from various data combinations.  The abbreviation ``W'' stands for WMAP, while ``P'' is Planck.
For one-sided intervals we show only the 95\%~credible limit, while for two-sided intervals we show the 95\% central credible interval (see~\cite{Hamann:2007pi} for definition) as well as the posterior mean.
The second last column compares the model's best-fit~$\chi^2$ value relative to the $\Lambda$CDM best-fit as per equation~(\ref{eq:chi2a}) in the face of $\Delta_\textrm{\tiny dof}$ 
additional degrees of freedom,
while the last column 
contrasts the chameleon and the $N_\textrm{\tiny eff}\Lambda$CDM fits as per equation~(\ref{eq:chi2b}).\label{tab2}}
\end{table}

Clearly, none of the data combinations prefer the chameleon model over the basic $\Lambda$CDM model.  This can be discerned firstly from the fact that only one-sided (instead of two-sided) limits exist for the chameleon parameters~$\alpha$ and~$\beta$, indicating that $\alpha=\beta=0$ are completely compatible with all data combinations.%
\footnote{Note that in practice, because we use equations~(\ref{eq:phi0}) and~(\ref{eq:fieldmass}) to determine the present-day field value~$\phi_0$ and the field mass~$M_\phi$, 
it is necessary to impose a nonzero lower limit on the priors on $\alpha$ and $\beta$, both chosen here to be $10^{ -5}$ as shown in table~\ref{tab1}, in order to avoid an artificial divergence at $\alpha= \beta=0$.
  However, because the phenomenology of $\alpha = \beta = 10^{-5}$ is for all purposes indistinguishable from $\Lambda$CDM,  we choose not to enforce the distinction between the two cases. }
Secondly, the different between the best-fit $\chi^2$ values of the chameleon and the $\Lambda$CDM model, 
\begin{equation}
\label{eq:chi2a}
\Delta \chi^2_\textrm{\tiny a}=\chi^2_{\textrm{\tiny model}}-\chi^2_{\Lambda \textrm{\tiny CDM}},
\end{equation}
is at best $-0.66$ for the Planck+WP+HighL+BAO data combination.   This is barely beyond the reliability limit of  $|\Delta \chi^2| \geq 0.6$ of the BOBYQA bounded minimisation routine~\cite{citeulike:5630011} used to find the best-fit $\chi^2$ values (the Metropolis--Hastings algorithm in standard MCMC analyses is not suitable for finding the maximum of a multi-dimensional likelihood function), and is in any case too small an improvement to the fit to warrant the introduction of two extra fit parameters.

\begin{figure}[t]
\begin{center}
{\resizebox{0.6\columnwidth}{!}{\includegraphics{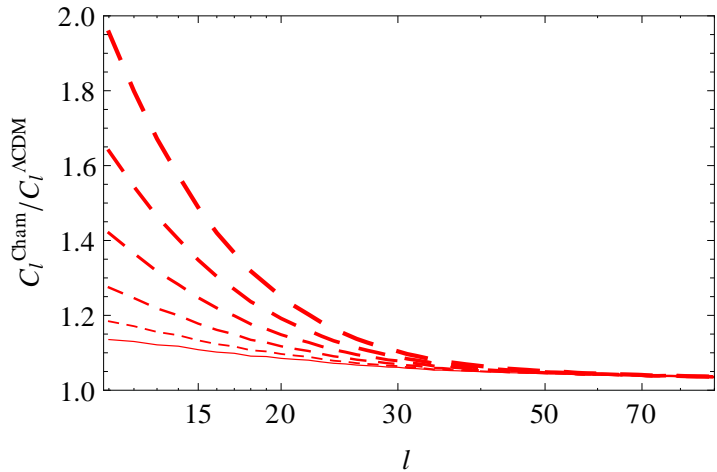}}}
\end{center}
\vspace{-0.6cm}
\caption{CMB TT spectrum of various chameleon models relative to the $\Lambda$CDM case.
The $\Lambda$CDM spectrum corresponds to the best-fit model of the data set Planck+WP+HST, while the chameleon models have the common parameters $(\alpha, \omega_\textrm{\tiny DM},h) = {(0.1, 0.117, 0.688)}$ and, from bottom to top, $\beta=0.05,0.1,0.15,0.2,0.25,0.3$.}\label{isw}
\end{figure}

In terms of the limits on the chameleon parameters~$\alpha$ and~$\beta$, there is virtually no difference between data combinations in their constraining power on the coupling parameter~$\beta$; all return the same 95\% upper limit $\beta < 0.19$.  This can be understood as follows: as already discussed in sections~\ref{sec:attractor} and~\ref{sec:cmb},
neither the effective dark energy equation of state~$w_\textrm{\tiny eff}$ nor the high-redshift observable~$z_\textrm{\tiny eq}$
is  sensitive to~$\beta$; the effect of $\beta$ is felt only through the late ISW effect, because of the enhanced Bardeen potential $\Phi_H$ at $z \lesssim 1$ on close to the Hubble length as shown in figure~\ref{figbardeen}.
Figure~\ref{isw} shows the CMB TT spectrum of various chameleon models relative to the $\Lambda$CDM best-fit to the Planck+WP+HST data combination.
 Although the $\beta$-dependence is clearly strong, the late ISW effect evidently manifests itself only at low multipoles $\ell \lesssim 30$.  This means firstly that, besides the CMB TT spectrum, 
the cosmological observations considered in this analysis play no  role in constraining $\beta$.  Secondly, swapping WMAP for Planck CMB temperature data, both of which are limited only by cosmic variance at the low multipoles, also cannot improve the sensitivity to $\beta$.  Further improvements can come about by either looking at the effects of the fifth force induced by $\beta$ on very nonlinear scales, or, in the future, the impact of the $\beta$-dependent metric perturbations on full-sky cosmic shear surveys such as the ESA Euclid mission~\cite{Refregier:2010ss}

\begin{figure}[t]
\begin{center}
{\resizebox{0.75\columnwidth}{!}{\includegraphics{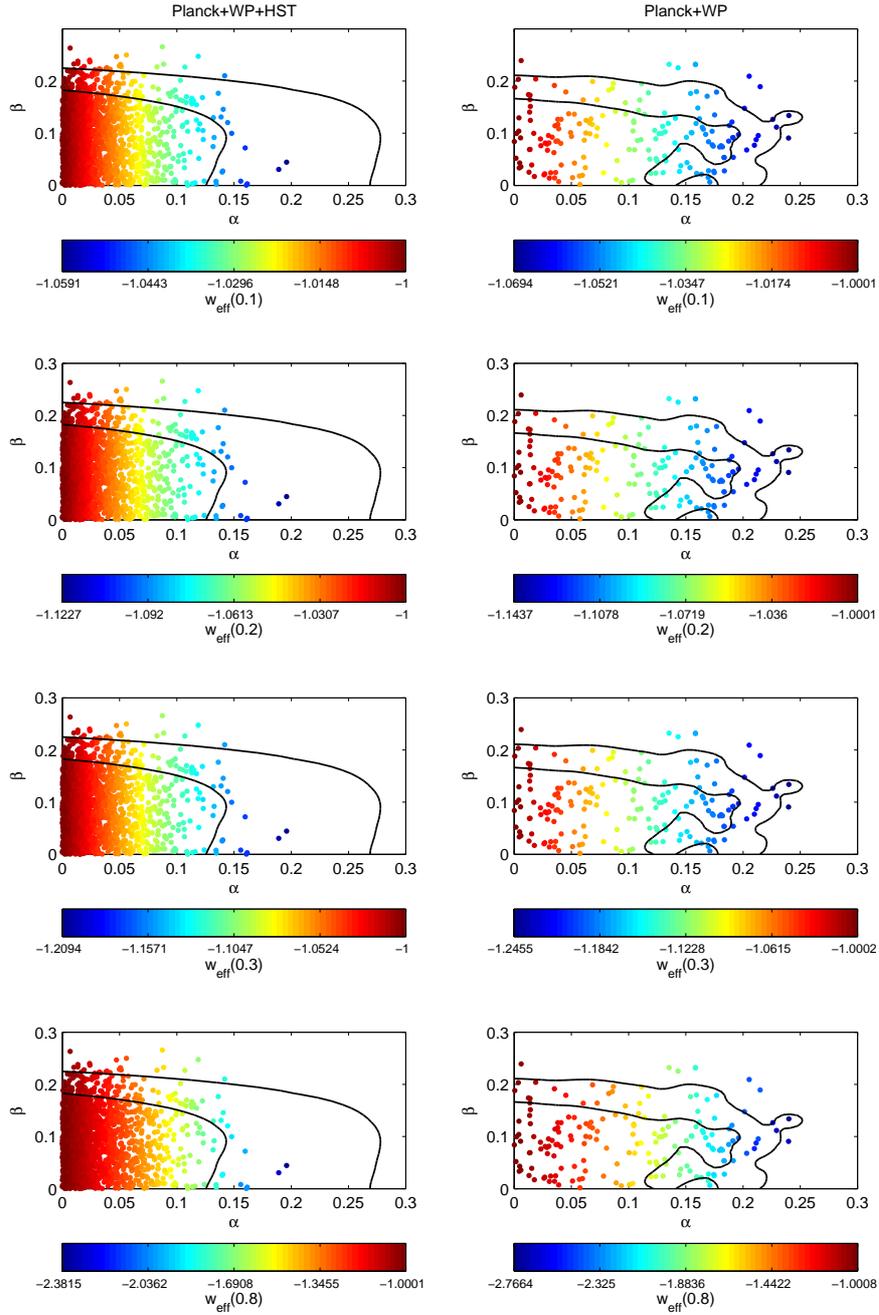}}}
\end{center}
\vspace{-0.6cm}
\caption{2D marginal 68\% and 95\% contours in the $(\alpha,\beta)$-plane derived from Planck+WP+HST (left) and Planck+WP (right), and the corresponding effective equation of state for the dark energy (coloured dots) at four different redshifts, from top to bottom,  $z=0.1, 0.2, 0.3, 0.8$.}\label{ab2}
\end{figure}

For the slope parameter~$\alpha$, the constraint generally improves when WMAP data are swapped for Planck measurements.  This can be understood as a consequence both of a more well-constrained $z_\textrm{\tiny eq}$ and of a more accurate determination of $\theta_\textrm{\tiny s}$.  The latter in particular constrains the effective equation of state~$w_\textrm{\tiny eff}$, whose departure from the canonical value of~$-1$ is now, as shown in figure~\ref{ab2} and table~\ref{tab3}, restricted to no more than 20\% (at 95\% confidence) at $z < 0.5$ and 50\% at $z < 0.8$.
Adding the HST measurement of~$H_0$ to Planck improves  the constraint on $\alpha$ even further (almost a factor of two).  However, this improvement may be an artefact originating from the $2\sigma+$ discrepancy between the HST measurement and the $h$~value inferred from Planck data in the minimal $\Lambda$CDM model~\cite{Ade:2013zuv,Planck:2015xua}.  Indeed, keeping $\omega_\textrm{\tiny DM}$ fixed, we see,
on the one hand,   from equation~(\ref{eq:zeq}) that  $\alpha$ must decrease with an increasing~$h$ in order  to maintain the same $z_\textrm{\tiny eq}$.  On the other hand, 
from equation~(\ref{eq:da}), maintaining the same $D_\textrm{\tiny A}^*$ in the face of a smaller $\alpha$ requires that we reduce $h$.  Thus, the net effect is that $\alpha$ goes down, while the inferred value of $h$ changes only marginally, as suggested by the numbers in table~\ref{tab2}.
This incompatibility between data sets is also  confirmed by the 2D marginal posterior in the $(\alpha, H_0)$-plane in figure~\ref{alphalikeli}, which shows a clear preference for a negative $\alpha$ value---which is not within the chameleon parameter space---if $H_0$ was indeed as large as $73.8\; {\rm km} \ {\rm s}^{-1} \ {\rm Mpc}^{-1}$ preferred by HST.

\begin{table}[t]
\begin{center}
{\footnotesize
 \begin{tabular}{|c|c|c|} \cline{1-3}
\multirow{2}{*}{$z$} 	& \multicolumn{2}{c|}{w$_\textrm{\tiny eff}(z)$}  \\ \cline{2-3}
	& Planck+WP+HST		    & Planck+WP           \\ \cline{1-3}
0.0 & $>-1.003$    & $>-1.005$  \\  \cline{1-3}
0.1 & $>-1.03$     & $>-1.06$   \\  \cline{1-3}
0.2 & $>-1.06$     & $>-1.12$   \\  \cline{1-3}
0.3 & $>-1.11$     & $>-1.21$   \\  \cline{1-3}
0.4 & $>-1.16$     & $>-1.32$   \\  \cline{1-3} 
0.5 & $>-1.2$      & $>-1.5$    \\  \cline{1-3}
0.6 & $>-1.3$      & $>-1.7$    \\  \cline{1-3}
0.7 & $>-1.4$      & $>-2.0$    \\  \cline{1-3}
0.8 & $>-1.5$      & $>-2.4$    \\  \cline{1-3}
\end{tabular}
}
\end{center}
\caption{1D marginal 95\% lower limit for the effective dark energy equation of state $w_{\rm eff}$ in the chameleon model at various redshifts, derived from the data combinations Planck+WP+HST and Planck+WP.\label{tab3}}
\end{table}

Comparing our results with previous investigations of interacting DM--DE models, our limits differ in significant ways.
 The analyses of~\cite{Das:2005yj,Corasaniti:2010ze} dealt with a scenario identical to ours, and it was argued in these works that~$\beta \sim  \mathcal{O}(1)$ could be compatible with observations.  However, this conclusion was based on the argument that with $\beta$ and $\alpha$ {\it fixed} at $1$ and $0.2$ respectively, a best-fit $\chi^2$ ``close to'' the $\Lambda$CDM best-fit could be obtained when the usual $\Lambda$CDM parameters are let to vary freely.  Our analysis here, in contrast, allows {\it all} parameters to vary, so as to locate the values of $\alpha$ and $\beta$ actually preferred by the observational data.
 References~\cite{LaVacca:2009yp} and~\cite{Pettorino:2013oxa} also considered a coupled scenario in which the scalar potential $V(\phi)$ is of the runaway form~(\ref{eq:pot}), and obtained constraints on $\beta$ that are considerably tighter than ours.  We note however that the coupling function adopted in these analyses is of the form $f(\phi) = e^{-\beta \phi/M_\textrm{\tiny Pl}}$, which differs from ours by a crucial minus sign, signifying that their effective potential~$V_\textrm{\tiny eff}(\phi)$ does not possess a local minimum, nor $\phi(t)$ an adiabatic attractor solution.  For this reason, the results of~\cite{LaVacca:2009yp} and~\cite{Pettorino:2013oxa} cannot be meaningfully compared with ours in the context of parameter inference.


\subsection{Chameleon mimicking dark radiation?} \label{sec:resulneff}

As discussed at the beginning of section~\ref{sec:lss}, the nonstandard time-evolution induced for the dark matter density by the DM--DE coupling
 tends to delay the epoch of matter--radiation equality relative to the standard $\Lambda$CDM case with the 
 same $\omega_\textrm{\tiny DM}$.  This raises the possibility that chameleon phenomenology might mimic the effect of dark radiation, which also has shifting~$z_\textrm{\tiny eff}$ as its primary effect.  We test this possibility by fitting the $\Lambda$CDM model extended with a free $N_{\textrm{\tiny eff}}$ to the same data combinations explored in 
section~\ref{sec:chamfits} for the chameleon model, and compute in each case the constraints on $N_\textrm{\tiny eff}$ as well as the best-fit $\chi^2$ value relative to $\Lambda$CDM as defined in equation~(\ref{eq:chi2a}).  In addition, for each data combination we evaluate the $\chi^2$~difference between the best-fit chameleon model and its
$N_\textrm{\tiny eff}\Lambda$CDM counterpart, defined as
\begin{equation}
\label{eq:chi2b}
\Delta \chi^2_\textrm{\tiny b}=\chi^2_{\textrm{\tiny chameleon}}-\chi^2_{N_\textrm{\tiny eff}\Lambda \textrm{\tiny CDM}}.
\end{equation}
This $\Delta \chi^2_\textrm{\tiny b}$ value will tell us whether or not chameleon models are able to mimic dark radiation.

Table~\ref{tab2} shows the results of this exercise.  Note that in deriving these constraints we have made use of the BBN consistency relation~\cite{Hamann:2007sb} in {\sc CosmoMC}, which automatically adjusts the helium mass fraction $Y_\textrm{\tiny He4}$ according to big bang nucleosynthesis predictions for each set of input parameters~$N_\textrm{\tiny eff}$ and~$\omega_\textrm{\tiny B}$.  Clearly, only the combinations WMAP+HST+SDSS and Planck+WP+HST show a $2 \sigma+$ preference for $N_\textrm{\tiny eff}> 3.046$, and in the latter case, this is accompanied by a dramatic increase in the inferred~$h$ value ($h= 0.73^{0.77}_{0.69}$) relative to other Planck data combinations.
In terms of the best-fit $\chi^2$, we find $\Delta \chi^2_\textrm{\tiny a}$ values of  $-0.98$ and $-4.32$ respectively relative to $\Lambda$CDM, suggesting a mild preference for a radiation excess especially in the latter case.

Comparing the chameleon model with $N_\textrm{\tiny eff}\Lambda$CDM for these same two data combinations, we find the chameleon scenario consistently the worse performer of the two.  The best-fit~$\chi^2$ differences between the two models as per definition~(\ref{eq:chi2b}) are  $+1.13$ and $+4.68$ for WMAP+HST+SDSS and Planck+WP+HST respectively.  This can be traced to the fact that phenomenologically, the $\alpha$ parameter needs to be negative in order to accommodate the large  $H_0$~value preferred by HST (see figure~{\ref{alphalikeli} and discussion in section~\ref{sec:chamfits}).   A negative~$\alpha$, however, does not belong to the chameleon parameter space.
Thus, despite a passing similarity at the equality epoch, the degeneracy between $H_0$ and the chameleon model parameter~$\alpha$  ultimately  works in the wrong direction to be able to resolve the tension between the HST measurement and the Planck-inferred $H_0$ value.

\begin{figure}[t]
\begin{center}
{\resizebox{0.50\columnwidth}{!}{\includegraphics{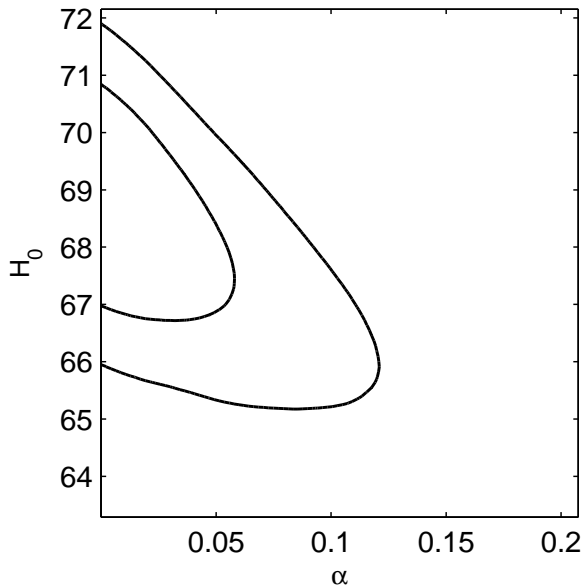}}}
\end{center}
\vspace{-0.6cm}
\caption{2D marginal posterior in the  $(\alpha, H_0)$-plane derived from Planck+WP+HST.}\label{alphalikeli}
\end{figure}

At this point, we note that our chameleon model is distinct from early dark energy models which {\it do} mimic a radiation excess because the dark energy equation of state in these models evolves to $\sim 1/3$ during radiation domination~\cite{Calabrese:2011hg}.   In contrast, the attractor solution in chameleon models ensures that the scalar field remains slowly rolling while sitting in the effective minimum of the potential, and therefore maintains a cosmological constant-like behaviour at all times.


\section{Future tests} \label{sec:prev}

The two main predictions of the chameleon model are the apparent phantom behaviour of the effective dark energy equation of state~$w_\textrm{\tiny eff}$ and a fifth force that enhances the clustering of dark matter as well as the gravitational potentials.  The former is, for any nonzero $\beta$, solely a function of the parameter~$\alpha$, while the latter is strongly dependent on $\beta$.

 As we have seen in figure~\ref{ab2} and table~\ref{tab3}, the effective dark energy equation of state is already constrained at 95\% confidence to $w_\textrm{\tiny eff} >-1.2$ at $z = 0.5$ and $<-1.5$ at $z=0.8$ by Planck+WP+HST.  It would therefore appear at first glance that the Dark Energy Survey (DES), with $1 \sigma$ sensitivities $\sigma(w_0) \sim 0.078$ and $\sigma(w_a)\sim 0.3$~\cite{Albrecht:2006um} for the parameterisation 
 \begin{equation}
 \label{eq:wa}
 w(a) = w_0 + w_a(1-a),
 \end{equation}
  may not be sufficient to further tighten the constraint on the chameleon parameter $\alpha$.  The ESA Euclid mission on the other hand, with $\sigma(w_0) \sim 0.018$ and $\sigma(w_a)\sim 0.15$~\cite{Refregier:2010ss}, should produce some improvements.  
  
We caution however that this interpretation is not strictly correct: parameter sensitivities derived for a simple dynamical dark model such as equation~(\ref{eq:wa}) do not automatically translate to bounds on the chameleon model parameters.  This is because ongoing and future large-scale structure surveys will derive their constraints on the dark energy equation of state from {\it both} distance measures and the growth of density perturbations;  apart from its effect on the late ISW effect, the latter observable has not been used in the present analysis of existing cosmological data.   Recall from our discussions in section~\ref{sec:cmb} that in the chameleon scenario, the evolution of the dark matter density parameter $\Omega_\textrm{\tiny DM}(t)$, equation~(\ref{eq:omegadmt}), which governs the gravitational potential felt by the density perturbations, maps to an effective dark energy equation of state {\it different}
from the canonical $w_\textrm{\tiny eff}$ responsible for the background Hubble expansion.  This additional feature should allow us to distinguish cleanly between simple dynamical dark energy and interacting/chameleon models.  A detailed analysis of how well future surveys will be able to select between dark energy models is beyond the scope of the present work, but we note that publicly available forecast codes such as~\cite{Hamann:2012fe,Basse:2013zua,Basse:2014qqa} exist which, with minor modifications, will be equal to the task.

As for the fifth force, there are two possible places to look for its effects.  One is the very nonlinear scales, where the perturbations have had the most time to evolve under the force's influence.  Given that the $\beta$ parameter is already constrained by the late ISW effect to less than $0.2$, we expect the fifth force to manifest itself on subhorizon scales, according to equation~(\ref{eq:deltadm2}), as a maximum 8\% correction to the (scale-dependent) effective Newtonian constant; nonlinear evolution should further enhance the effect (although difficult to predict precisely).
The second possibility is to take advantage of the almost full sky coverage of future surveys such as the ESA Euclid mission.  In the same way that the scale-dependent evolution of the Bardeen potential $\Phi_H$ at $z \lesssim 1$ and $k \sim (a H)^{-1}$ induced by $\beta$ (see figure~\ref{figbardeen}) enhances the late ISW effect, we expect an analogous distortion of, 
e.g., the cosmic shear convergence angular power spectrum, at similarly low multipoles $\ell \lesssim 30$.  Of course, detection of this signature and ultimately the extent to which one could improve upon  existing constraints on~$\beta$ will still be, like the late ISW effect, limited by cosmic variance.  To address this question quantitatively we would need to perform a parameter sensitivity forecast, which we defer to a future work.


\section{Conclusions} \label{sec:conclusions}

We have considered in this work a simple interacting DM--DE/chameleon model in which a scalar field slowly rolls and adiabatically tracks the minimum of an effective potential starting from deep in the radiation-dominated era.   We have tested this model against recent cosmological observations, especially new CMB temperature measurements from the Planck mission, and have found constraints on the two model parameters: $\alpha$, slope of the scalar potential, and $\beta$, the coupling strength.
We find that while the constraint on the coupling strength $\beta$ has remained unchanged between WMAP and Planck, essentially because this parameter is constrained primarily by the late ISW effect, the upper limit on $\alpha$, which has the dual effect of shifting the matter--radiation equality epoch and altering the effective dark energy equation of state, has improved somewhat.

Because of its potential to shift the epoch of matter--radiation equality, we have tested also the possibility that such a chameleon model might mimic the phenomenology of dark radiation, and resolve the apparent tension between the relatively large Hubble expansion rate measured in our local neighbourhood by HST and the low value inferred from the Planck CMB data.  We find however that the degeneracy between $\alpha$ and $H_0$ in the chameleon model works in the wrong direction: a larger $H_0$ value in tends to push $\alpha$ towards zero, i.e., the cosmological constant limit.  In fact, the ultimate effect of the HST/Planck tension on the chameleon model is to put an artificially tight constraint on $\alpha$.

Taking into account the allowed model parameter space, we have discussed possible signatures of this type of chameleon model for ongoing and future large-scale structure surveys such as DES~\cite{Albrecht:2006um} and the ESA Euclid mission~\cite{Refregier:2010ss}.  The salient effects are (i) a mismatch between the effective dark energy equation of state inferred from the background expansion and from the growth function which should allow us to determine or further constrain the value of~$\alpha$, and (ii) distortion of the large-scale distribution power spectrum on very nonlinear and close-to-horizon scales, both of which are predominantly sensitive to $\beta$.
To determine the precise sensitivities of DES or Euclid to the chameleon model parameters would however require a dedicated forecast analysis.  We leave this exercise for a future work.

\section*{Acknowledgments}

DB thanks the Institut f\"ur Theoretische Teilchenphysik und Kosmologie at RWTH Aachen for their hospitality during the development of this work. DB acknowledges financial support from CAPES-PDEE (grant 6864-10-4), and the use of computing resources at CENAPAD-SP/UNICAMP and CCJDR-IFGW/UNICAMP.

\bibliographystyle{utcaps}

\bibliography{newrefs}

\providecommand{\href}[2]{#2}\begingroup\raggedright\begin{thebibliography}{10}

\bibitem{Riess:1998cb}
{\bfseries Supernova Search Team} Collaboration, A.~G. Riess {\em et al.},
  ``{Observational evidence from supernovae for an accelerating universe and a
  cosmological constant},'' \href{http://dx.doi.org/10.1086/300499}{{\em
  Astron.J.} {\bfseries 116} (1998)  1009--1038},
\href{http://arxiv.org/abs/astro-ph/9805201}{{\ttfamily arXiv:astro-ph/9805201
  [astro-ph]}}.

\bibitem{Perlmutter:1998np}
{\bfseries Supernova Cosmology Project} Collaboration, S.~Perlmutter {\em et
  al.}, ``{Measurements of Omega and Lambda from 42 high redshift
  supernovae},'' \href{http://dx.doi.org/10.1086/307221}{{\em Astrophys.J.}
  {\bfseries 517} (1999)  565--586},
\href{http://arxiv.org/abs/astro-ph/9812133}{{\ttfamily arXiv:astro-ph/9812133
  [astro-ph]}}.

\bibitem{Suzuki:2011hu}
N.~Suzuki, D.~Rubin, C.~Lidman, G.~Aldering, R.~Amanullah, {\em et al.}, ``{The
  Hubble Space Telescope Cluster Supernova Survey: V. Improving the Dark Energy
  Constraints Above z<1 and Building an Early-Type-Hosted Supernova Sample},''
  \href{http://dx.doi.org/10.1088/0004-637X/746/1/85}{{\em Astrophys.J.}
  {\bfseries 746} (2012)  85},
\href{http://arxiv.org/abs/1105.3470}{{\ttfamily arXiv:1105.3470
  [astro-ph.CO]}}.

\bibitem{Sherwin:2011gv}
B.~D. Sherwin, J.~Dunkley, S.~Das, J.~W. Appel, J.~R. Bond, {\em et al.},
  ``{Evidence for dark energy from the cosmic microwave background alone using
  the Atacama Cosmology Telescope lensing measurements},''
  \href{http://dx.doi.org/10.1103/PhysRevLett.107.021302}{{\em Phys.Rev.Lett.}
  {\bfseries 107} (2011)  021302},
\href{http://arxiv.org/abs/1105.0419}{{\ttfamily arXiv:1105.0419
  [astro-ph.CO]}}.

\bibitem{Ade:2013zuv}
{\bfseries Planck} Collaboration, P.~Ade {\em et al.}, ``{Planck 2013 results.
  XVI. Cosmological parameters},''
\href{http://arxiv.org/abs/1303.5076}{{\ttfamily arXiv:1303.5076
  [astro-ph.CO]}}.

\bibitem{Planck:2015xua}
{\bfseries Planck} Collaboration, P.~Ade {\em et al.}, ``{Planck 2015 results.
  XIII. Cosmological parameters},''
\href{http://arxiv.org/abs/1502.01589}{{\ttfamily arXiv:1502.01589
  [astro-ph.CO]}}.

\bibitem{Wetterich:1987fm}
C.~Wetterich, ``{Cosmology and the Fate of Dilatation Symmetry},''
\href{http://dx.doi.org/10.1016/0550-3213(88)90193-9}{{\em Nucl.Phys.}
  {\bfseries B302} (1988)  668}.

\bibitem{Ratra:1987rm}
B.~Ratra and P.~Peebles, ``{Cosmological Consequences of a Rolling Homogeneous
  Scalar Field},''
\href{http://dx.doi.org/10.1103/PhysRevD.37.3406}{{\em Phys.Rev.} {\bfseries
  D37} (1988)  3406}.

\bibitem{Caldwell:1997ii}
R.~Caldwell, R.~Dave, and P.~J. Steinhardt, ``{Cosmological imprint of an
  energy component with general equation of state},''
  \href{http://dx.doi.org/10.1103/PhysRevLett.80.1582}{{\em Phys.Rev.Lett.}
  {\bfseries 80} (1998)  1582--1585},
\href{http://arxiv.org/abs/astro-ph/9708069}{{\ttfamily arXiv:astro-ph/9708069
  [astro-ph]}}.

\bibitem{Lombriser:2010mp}
L.~Lombriser, A.~Slosar, U.~Seljak, and W.~Hu, ``{Constraints on f(R) gravity
  from probing the large-scale structure},''
  \href{http://dx.doi.org/10.1103/PhysRevD.85.124038}{{\em Phys.Rev.}
  {\bfseries D85} (2012)  124038},
\href{http://arxiv.org/abs/1003.3009}{{\ttfamily arXiv:1003.3009
  [astro-ph.CO]}}.

\bibitem{Dossett:2014oia}
J.~Dossett, B.~Hu, and D.~Parkinson, ``{Constraining models of f(R) gravity
  with Planck and WiggleZ power spectrum data},''
  \href{http://dx.doi.org/10.1088/1475-7516/2014/03/046}{{\em JCAP} {\bfseries
  1403} (2014)  046},
\href{http://arxiv.org/abs/1401.3980}{{\ttfamily arXiv:1401.3980
  [astro-ph.CO]}}.

\bibitem{Bel:2014awa}
J.~Bel, P.~Brax, C.~Marinoni, and P.~Valageas, ``{Cosmological tests of
  modified gravity: constraints on $F(R)$ theories from the galaxy clustering
  ratio},''
\href{http://arxiv.org/abs/1406.3347}{{\ttfamily arXiv:1406.3347
  [astro-ph.CO]}}.

\bibitem{Amendola:1999er}
L.~Amendola, ``{Coupled quintessence},''
  \href{http://dx.doi.org/10.1103/PhysRevD.62.043511}{{\em Phys.Rev.}
  {\bfseries D62} (2000)  043511},
\href{http://arxiv.org/abs/astro-ph/9908023}{{\ttfamily arXiv:astro-ph/9908023
  [astro-ph]}}.

\bibitem{Faulkner:2006ub}
T.~Faulkner, M.~Tegmark, E.~F. Bunn, and Y.~Mao, ``{Constraining f(R) Gravity
  as a Scalar Tensor Theory},''
  \href{http://dx.doi.org/10.1103/PhysRevD.76.063505}{{\em Phys.Rev.}
  {\bfseries D76} (2007)  063505},
\href{http://arxiv.org/abs/astro-ph/0612569}{{\ttfamily arXiv:astro-ph/0612569
  [astro-ph]}}.

\bibitem{Hu:2007pj}
W.~Hu and I.~Sawicki, ``{A Parameterized Post-Friedmann Framework for Modified
  Gravity},'' \href{http://dx.doi.org/10.1103/PhysRevD.76.104043}{{\em
  Phys.Rev.} {\bfseries D76} (2007)  104043},
\href{http://arxiv.org/abs/0708.1190}{{\ttfamily arXiv:0708.1190 [astro-ph]}}.

\bibitem{Bertschinger:2008zb}
E.~Bertschinger and P.~Zukin, ``{Distinguishing Modified Gravity from Dark
  Energy},'' \href{http://dx.doi.org/10.1103/PhysRevD.78.024015}{{\em
  Phys.Rev.} {\bfseries D78} (2008)  024015},
\href{http://arxiv.org/abs/0801.2431}{{\ttfamily arXiv:0801.2431 [astro-ph]}}.

\bibitem{Chimento:2003iea}
L.~P. Chimento, A.~S. Jakubi, D.~Pavon, and W.~Zimdahl, ``{Interacting
  quintessence solution to the coincidence problem},''
  \href{http://dx.doi.org/10.1103/PhysRevD.67.083513}{{\em Phys.Rev.}
  {\bfseries D67} (2003)  083513},
\href{http://arxiv.org/abs/astro-ph/0303145}{{\ttfamily arXiv:astro-ph/0303145
  [astro-ph]}}.

\bibitem{Comelli:2003cv}
D.~Comelli, M.~Pietroni, and A.~Riotto, ``{Dark energy and dark matter},''
  \href{http://dx.doi.org/10.1016/j.physletb.2003.05.006}{{\em Phys.Lett.}
  {\bfseries B571} (2003)  115--120},
\href{http://arxiv.org/abs/hep-ph/0302080}{{\ttfamily arXiv:hep-ph/0302080
  [hep-ph]}}.

\bibitem{Will:2001mx}
C.~M. Will, ``{The Confrontation between general relativity and experiment},''
  {\em Living Rev.Rel.} {\bfseries 4} (2001)  4,
\href{http://arxiv.org/abs/gr-qc/0103036}{{\ttfamily arXiv:gr-qc/0103036
  [gr-qc]}}.

\bibitem{Kapner:2006si}
D.~Kapner, T.~Cook, E.~Adelberger, J.~Gundlach, B.~R. Heckel, {\em et al.},
  ``{Tests of the gravitational inverse-square law below the dark-energy length
  scale},'' \href{http://dx.doi.org/10.1103/PhysRevLett.98.021101}{{\em
  Phys.Rev.Lett.} {\bfseries 98} (2007)  021101},
\href{http://arxiv.org/abs/hep-ph/0611184}{{\ttfamily arXiv:hep-ph/0611184
  [hep-ph]}}.

\bibitem{Hoyle:2004cw}
C.~Hoyle, D.~Kapner, B.~R. Heckel, E.~Adelberger, J.~Gundlach, {\em et al.},
  ``{Sub-millimeter tests of the gravitational inverse-square law},''
  \href{http://dx.doi.org/10.1103/PhysRevD.70.042004}{{\em Phys.Rev.}
  {\bfseries D70} (2004)  042004},
\href{http://arxiv.org/abs/hep-ph/0405262}{{\ttfamily arXiv:hep-ph/0405262
  [hep-ph]}}.

\bibitem{Yang:2012zzb}
S.-Q. Yang, B.-F. Zhan, Q.-L. Wang, C.-G. Shao, L.-C. Tu, {\em et al.}, ``{Test
  of the Gravitational Inverse Square Law at Millimeter Ranges},''
\href{http://dx.doi.org/10.1103/PhysRevLett.108.081101}{{\em Phys.Rev.Lett.}
  {\bfseries 108} (2012)  081101}.

\bibitem{Khoury:2003aq}
J.~Khoury and A.~Weltman, ``{Chameleon fields: Awaiting surprises for tests of
  gravity in space},''
  \href{http://dx.doi.org/10.1103/PhysRevLett.93.171104}{{\em Phys.Rev.Lett.}
  {\bfseries 93} (2004)  171104},
\href{http://arxiv.org/abs/astro-ph/0309300}{{\ttfamily arXiv:astro-ph/0309300
  [astro-ph]}}.

\bibitem{Khoury:2003rn}
J.~Khoury and A.~Weltman, ``{Chameleon cosmology},''
  \href{http://dx.doi.org/10.1103/PhysRevD.69.044026}{{\em Phys.Rev.}
  {\bfseries D69} (2004)  044026},
\href{http://arxiv.org/abs/astro-ph/0309411}{{\ttfamily arXiv:astro-ph/0309411
  [astro-ph]}}.

\bibitem{Brax:2004qh}
P.~Brax, C.~van~de Bruck, A.-C. Davis, J.~Khoury, and A.~Weltman, ``{Detecting
  dark energy in orbit - The Cosmological chameleon},''
  \href{http://dx.doi.org/10.1103/PhysRevD.70.123518}{{\em Phys.Rev.}
  {\bfseries D70} (2004)  123518},
\href{http://arxiv.org/abs/astro-ph/0408415}{{\ttfamily arXiv:astro-ph/0408415
  [astro-ph]}}.

\bibitem{Brax:2005ew}
P.~Brax, C.~van~de Bruck, A.-C. Davis, and A.~M. Green, ``{Small scale
  structure formation in chameleon cosmology},''
  \href{http://dx.doi.org/10.1016/j.physletb.2005.12.055}{{\em Phys.Lett.}
  {\bfseries B633} (2006)  441--446},
\href{http://arxiv.org/abs/astro-ph/0509878}{{\ttfamily arXiv:astro-ph/0509878
  [astro-ph]}}.

\bibitem{Copeland:2006wr}
E.~J. Copeland, M.~Sami, and S.~Tsujikawa, ``{Dynamics of dark energy},''
  \href{http://dx.doi.org/10.1142/S021827180600942X}{{\em Int.J.Mod.Phys.}
  {\bfseries D15} (2006)  1753--1936},
\href{http://arxiv.org/abs/hep-th/0603057}{{\ttfamily arXiv:hep-th/0603057
  [hep-th]}}.

\bibitem{2012PhRvL.109x1301W}
J.~Wang, L.~Hui, and J.~Khoury, ``{No-Go Theorems for Generalized Chameleon
  Field Theories},''
  \href{http://dx.doi.org/{10.1103/PhysRevLett.109.241301}}{{\em {Physical
  Review Letters}} {\bfseries 109} (2012) no.~24, },
  \href{http://arxiv.org/abs/{1208.4612}}{{\ttfamily arXiv:{1208.4612}
  [astro-ph.CO]}}.

\bibitem{Das:2005yj}
S.~Das, P.~S. Corasaniti, and J.~Khoury, ``{Super-acceleration as signature of
  dark sector interaction},''
  \href{http://dx.doi.org/10.1103/PhysRevD.73.083509}{{\em Phys.Rev.}
  {\bfseries D73} (2006)  083509},
\href{http://arxiv.org/abs/astro-ph/0510628}{{\ttfamily arXiv:astro-ph/0510628
  [astro-ph]}}.

\bibitem{Damour:1994zq}
T.~Damour and A.~M. Polyakov, ``{The String dilaton and a least coupling
  principle},'' \href{http://dx.doi.org/10.1016/0550-3213(94)90143-0}{{\em
  Nucl.Phys.} {\bfseries B423} (1994)  532--558},
\href{http://arxiv.org/abs/hep-th/9401069}{{\ttfamily arXiv:hep-th/9401069
  [hep-th]}}.

\bibitem{Dunkley:2010ge}
J.~Dunkley, R.~Hlozek, J.~Sievers, V.~Acquaviva, P.~Ade, {\em et al.}, ``{The
  Atacama Cosmology Telescope: Cosmological Parameters from the 2008 Power
  Spectra},'' \href{http://dx.doi.org/10.1088/0004-637X/739/1/52}{{\em
  Astrophys.J.} {\bfseries 739} (2011)  52},
\href{http://arxiv.org/abs/1009.0866}{{\ttfamily arXiv:1009.0866
  [astro-ph.CO]}}.

\bibitem{Keisler:2011aw}
R.~Keisler, C.~Reichardt, K.~Aird, B.~Benson, L.~Bleem, {\em et al.}, ``{A
  Measurement of the Damping Tail of the Cosmic Microwave Background Power
  Spectrum with the South Pole Telescope},''
  \href{http://dx.doi.org/10.1088/0004-637X/743/1/28}{{\em Astrophys.J.}
  {\bfseries 743} (2011)  28},
\href{http://arxiv.org/abs/1105.3182}{{\ttfamily arXiv:1105.3182
  [astro-ph.CO]}}.

\bibitem{Hamann:2010bk}
J.~Hamann, S.~Hannestad, G.~G. Raffelt, I.~Tamborra, and Y.~Y. Wong,
  ``{Cosmology seeking friendship with sterile neutrinos},''
  \href{http://dx.doi.org/10.1103/PhysRevLett.105.181301}{{\em Phys.Rev.Lett.}
  {\bfseries 105} (2010)  181301},
\href{http://arxiv.org/abs/1006.5276}{{\ttfamily arXiv:1006.5276 [hep-ph]}}.

\bibitem{Hamann:2011ge}
J.~Hamann, S.~Hannestad, G.~G. Raffelt, and Y.~Y. Wong, ``{Sterile neutrinos
  with eV masses in cosmology: How disfavoured exactly?},''
  \href{http://dx.doi.org/10.1088/1475-7516/2011/09/034}{{\em JCAP} {\bfseries
  1109} (2011)  034},
\href{http://arxiv.org/abs/1108.4136}{{\ttfamily arXiv:1108.4136
  [astro-ph.CO]}}.

\bibitem{Vogel:2013raa}
H.~Vogel and J.~Redondo, ``{Dark Radiation constraints on minicharged particles
  in models with a hidden photon},''
  \href{http://dx.doi.org/10.1088/1475-7516/2014/02/029}{{\em JCAP} {\bfseries
  1402} (2014)  029},
\href{http://arxiv.org/abs/1311.2600}{{\ttfamily arXiv:1311.2600 [hep-ph]}}.

\bibitem{Hasenkamp:2014hma}
J.~Hasenkamp, ``{Daughters mimic sterile neutrinos (almost!) perfectly},''
\href{http://arxiv.org/abs/1405.6736}{{\ttfamily arXiv:1405.6736
  [astro-ph.CO]}}.

\bibitem{Bjaelde:2010vt}
O.~E. Bjaelde and S.~Das, ``{Dark Matter Decaying into a Fermi Sea of
  Neutrinos},'' \href{http://dx.doi.org/10.1103/PhysRevD.82.043504}{{\em
  Phys.Rev.} {\bfseries D82} (2010)  043504},
\href{http://arxiv.org/abs/1002.1306}{{\ttfamily arXiv:1002.1306
  [astro-ph.CO]}}.

\bibitem{DiBari:2013dna}
P.~Di~Bari, S.~F. King, and A.~Merle, ``{Dark Radiation or Warm Dark Matter
  from long lived particle decays in the light of Planck},''
  \href{http://dx.doi.org/10.1016/j.physletb.2013.06.003}{{\em Phys.Lett.}
  {\bfseries B724} (2013)  77--83},
\href{http://arxiv.org/abs/1303.6267}{{\ttfamily arXiv:1303.6267 [hep-ph]}}.

\bibitem{Linder:2008nq}
E.~V. Linder and G.~Robbers, ``{Shifting the Universe: Early Dark Energy and
  Standard Rulers},''
  \href{http://dx.doi.org/10.1088/1475-7516/2008/06/004}{{\em JCAP} {\bfseries
  0806} (2008)  004},
\href{http://arxiv.org/abs/0803.2877}{{\ttfamily arXiv:0803.2877 [astro-ph]}}.

\bibitem{Hou:2011ec}
Z.~Hou, R.~Keisler, L.~Knox, M.~Millea, and C.~Reichardt, ``{How Massless
  Neutrinos Affect the Cosmic Microwave Background Damping Tail},''
  \href{http://dx.doi.org/10.1103/PhysRevD.87.083008}{{\em Phys.Rev.}
  {\bfseries D87} (2013)  083008},
\href{http://arxiv.org/abs/1104.2333}{{\ttfamily arXiv:1104.2333
  [astro-ph.CO]}}.

\bibitem{Archidiacono:2013cha}
M.~Archidiacono, S.~Hannestad, A.~Mirizzi, G.~Raffelt, and Y.~Y. Wong, ``{Axion
  hot dark matter bounds after Planck},''
  \href{http://dx.doi.org/10.1088/1475-7516/2013/10/020}{{\em JCAP} {\bfseries
  1310} (2013)  020},
\href{http://arxiv.org/abs/1307.0615}{{\ttfamily arXiv:1307.0615
  [astro-ph.CO]}}.

\bibitem{Lewis:1999bs}
A.~Lewis, A.~Challinor, and A.~Lasenby, ``{Efficient computation of CMB
  anisotropies in closed FRW models},''
  \href{http://dx.doi.org/10.1086/309179}{{\em Astrophys.J.} {\bfseries 538}
  (2000)  473--476},
\href{http://arxiv.org/abs/astro-ph/9911177}{{\ttfamily arXiv:astro-ph/9911177
  [astro-ph]}}.

\bibitem{Bean:2007ny}
R.~Bean, E.~E. Flanagan, and M.~Trodden, ``{Adiabatic instability in coupled
  dark energy-dark matter models},''
  \href{http://dx.doi.org/10.1103/PhysRevD.78.023009}{{\em Phys.Rev.}
  {\bfseries D78} (2008)  023009},
\href{http://arxiv.org/abs/0709.1128}{{\ttfamily arXiv:0709.1128 [astro-ph]}}.

\bibitem{Corasaniti:2008kx}
P.~S. Corasaniti, ``{Slow-Roll Suppression of Adiabatic Instabilities in
  Coupled Scalar Field-Dark Matter Models},''
  \href{http://dx.doi.org/10.1103/PhysRevD.78.083538}{{\em Phys.Rev.}
  {\bfseries D78} (2008)  083538},
\href{http://arxiv.org/abs/0808.1646}{{\ttfamily arXiv:0808.1646 [astro-ph]}}.

\bibitem{Reid:2009xm}
B.~A. Reid, W.~J. Percival, D.~J. Eisenstein, L.~Verde, D.~N. Spergel, {\em et
  al.}, ``{Cosmological Constraints from the Clustering of the Sloan Digital
  Sky Survey DR7 Luminous Red Galaxies},''
  \href{http://dx.doi.org/10.1111/j.1365-2966.2010.16276.x}{{\em
  Mon.Not.Roy.Astron.Soc.} {\bfseries 404} (2010)  60--85},
\href{http://arxiv.org/abs/0907.1659}{{\ttfamily arXiv:0907.1659
  [astro-ph.CO]}}.

\bibitem{Motohashi:2012wc}
H.~Motohashi, A.~A. Starobinsky, and J.~Yokoyama, ``{Cosmology Based on f(R)
  Gravity Admits 1 eV Sterile Neutrinos},''
  \href{http://dx.doi.org/10.1103/PhysRevLett.110.121302}{{\em Phys.Rev.Lett.}
  {\bfseries 110} (2013) no.~12, 121302},
\href{http://arxiv.org/abs/1203.6828}{{\ttfamily arXiv:1203.6828
  [astro-ph.CO]}}.

\bibitem{Brax:2011ta}
P.~Brax and A.-C. Davis, ``{Modified Gravity and the CMB},''
  \href{http://dx.doi.org/10.1103/PhysRevD.85.023513}{{\em Phys.Rev.}
  {\bfseries D85} (2012)  023513},
\href{http://arxiv.org/abs/1109.5862}{{\ttfamily arXiv:1109.5862
  [astro-ph.CO]}}.

\bibitem{Reichardt:2012yj}
C.~Reichardt, B.~Stalder, L.~Bleem, T.~Montroy, K.~Aird, {\em et al.},
  ``{Galaxy clusters discovered via the Sunyaev-Zel'dovich effect in the first
  720 square degrees of the South Pole Telescope survey},''
  \href{http://dx.doi.org/10.1088/0004-637X/763/2/127}{{\em Astrophys.J.}
  {\bfseries 763} (2013)  127},
\href{http://arxiv.org/abs/1203.5775}{{\ttfamily arXiv:1203.5775
  [astro-ph.CO]}}.

\bibitem{Ade:2013skr}
{\bfseries Planck} Collaboration, P.~Ade {\em et al.}, ``{Planck 2013 results.
  XXIX. Planck catalogue of Sunyaev-Zeldovich sources},'' {\em
  Astron.Astrophys.} (2013)  ,
\href{http://arxiv.org/abs/1303.5089}{{\ttfamily arXiv:1303.5089
  [astro-ph.CO]}}.

\bibitem{Mantz:2014paa}
A.~B. Mantz, A.~von~der Linden, S.~W. Allen, D.~E. Applegate, P.~L. Kelly, {\em
  et al.}, ``{Weighing the Giants IV: Cosmology and Neutrino Mass},''
\href{http://arxiv.org/abs/1407.4516}{{\ttfamily arXiv:1407.4516
  [astro-ph.CO]}}.

\bibitem{Schrabback:2009ba}
T.~Schrabback, J.~Hartlap, B.~Joachimi, M.~Kilbinger, P.~Simon, {\em et al.},
  ``{Evidence for the accelerated expansion of the Universe from weak lensing
  tomography with COSMOS},''
  \href{http://dx.doi.org/10.1051/0004-6361/200913577}{{\em Astron.Astrophys.}
  {\bfseries 516} (2010)  A63},
\href{http://arxiv.org/abs/0911.0053}{{\ttfamily arXiv:0911.0053
  [astro-ph.CO]}}.

\bibitem{Jee:2012hr}
M.~J. Jee, J.~A. Tyson, M.~D. Schneider, D.~Wittman, S.~Schmidt, {\em et al.},
  ``{Cosmic shear results from the deep lens survey - I: Joint constraints on
  $\omega_m$ and $\sigma_8$ with a two-dimensional analysis},''
  \href{http://dx.doi.org/10.1088/0004-637X/765/1/74}{{\em Astrophys.J.}
  {\bfseries 765} (2013)  74},
\href{http://arxiv.org/abs/1210.2732}{{\ttfamily arXiv:1210.2732
  [astro-ph.CO]}}.

\bibitem{Kitching:2014dtq}
{\bfseries CFHTLenS} Collaboration, T.~Kitching {\em et al.}, ``{3D Cosmic
  Shear: Cosmology from CFHTLenS},''
\href{http://arxiv.org/abs/1401.6842}{{\ttfamily arXiv:1401.6842
  [astro-ph.CO]}}.

\bibitem{McDonald:2004eu}
{\bfseries SDSS} Collaboration, P.~McDonald {\em et al.}, ``{The Lyman-alpha
  forest power spectrum from the Sloan Digital Sky Survey},''
  \href{http://dx.doi.org/10.1086/444361}{{\em Astrophys.J.Suppl.} {\bfseries
  163} (2006)  80--109},
\href{http://arxiv.org/abs/astro-ph/0405013}{{\ttfamily arXiv:astro-ph/0405013
  [astro-ph]}}.

\bibitem{Lee:2012xb}
K.-G. Lee, S.~Bailey, L.~E. Bartsch, W.~Carithers, K.~S. Dawson, {\em et al.},
  ``{The BOSS Lyman-alpha Forest Sample from SDSS Data Release 9},''
\href{http://arxiv.org/abs/1211.5146}{{\ttfamily arXiv:1211.5146
  [astro-ph.CO]}}.

\bibitem{Lewis:2002ah}
A.~Lewis and S.~Bridle, ``{Cosmological parameters from CMB and other data: A
  Monte Carlo approach},''
  \href{http://dx.doi.org/10.1103/PhysRevD.66.103511}{{\em Phys.Rev.}
  {\bfseries D66} (2002)  103511},
\href{http://arxiv.org/abs/astro-ph/0205436}{{\ttfamily arXiv:astro-ph/0205436
  [astro-ph]}}.

\bibitem{Hinshaw:2012aka}
{\bfseries WMAP} Collaboration, G.~Hinshaw {\em et al.}, ``{Nine-Year Wilkinson
  Microwave Anisotropy Probe (WMAP) Observations: Cosmological Parameter
  Results},'' \href{http://dx.doi.org/10.1088/0067-0049/208/2/19}{{\em
  Astrophys.J.Suppl.} {\bfseries 208} (2013)  19},
\href{http://arxiv.org/abs/1212.5226}{{\ttfamily arXiv:1212.5226
  [astro-ph.CO]}}.

\bibitem{Dunkley:2013vu}
J.~Dunkley, E.~Calabrese, J.~Sievers, G.~Addison, N.~Battaglia, {\em et al.},
  ``{The Atacama Cosmology Telescope: likelihood for small-scale CMB data},''
\href{http://arxiv.org/abs/1301.0776}{{\ttfamily arXiv:1301.0776
  [astro-ph.CO]}}.

\bibitem{Das:2013zf}
S.~Das, T.~Louis, M.~R. Nolta, G.~E. Addison, E.~S. Battistelli, {\em et al.},
  ``{The Atacama Cosmology Telescope: temperature and gravitational lensing
  power spectrum measurements from three seasons of data},''
  \href{http://dx.doi.org/10.1088/1475-7516/2014/04/014}{{\em JCAP} {\bfseries
  1404} (2014)  014},
\href{http://arxiv.org/abs/1301.1037}{{\ttfamily arXiv:1301.1037
  [astro-ph.CO]}}.

\bibitem{Story:2012wx}
K.~Story, C.~Reichardt, Z.~Hou, R.~Keisler, K.~Aird, {\em et al.}, ``{A
  Measurement of the Cosmic Microwave Background Damping Tail from the
  2500-square-degree SPT-SZ survey},''
  \href{http://dx.doi.org/10.1088/0004-637X/779/1/86}{{\em Astrophys.J.}
  {\bfseries 779} (2013)  86},
\href{http://arxiv.org/abs/1210.7231}{{\ttfamily arXiv:1210.7231
  [astro-ph.CO]}}.

\bibitem{Hou:2012xq}
Z.~Hou, C.~Reichardt, K.~Story, B.~Follin, R.~Keisler, {\em et al.},
  ``{Constraints on Cosmology from the Cosmic Microwave Background Power
  Spectrum of the 2500-square degree SPT-SZ Survey},''
  \href{http://dx.doi.org/10.1088/0004-637X/782/2/74}{{\em Astrophys.J.}
  {\bfseries 782} (2014)  74},
\href{http://arxiv.org/abs/1212.6267}{{\ttfamily arXiv:1212.6267
  [astro-ph.CO]}}.

\bibitem{Calabrese:2013jyk}
E.~Calabrese, R.~A. Hlozek, N.~Battaglia, E.~S. Battistelli, J.~R. Bond, {\em
  et al.}, ``{Cosmological parameters from pre-planck cosmic microwave
  background measurements},''
  \href{http://dx.doi.org/10.1103/PhysRevD.87.103012}{{\em Phys.Rev.}
  {\bfseries D87} (2013) no.~10, 103012},
\href{http://arxiv.org/abs/1302.1841}{{\ttfamily arXiv:1302.1841
  [astro-ph.CO]}}.

\bibitem{Riess:2009pu}
A.~G. Riess, L.~Macri, S.~Casertano, M.~Sosey, H.~Lampeitl, {\em et al.}, ``{A
  Redetermination of the Hubble Constant with the Hubble Space Telescope from a
  Differential Distance Ladder},''
  \href{http://dx.doi.org/10.1088/0004-637X/699/1/539}{{\em Astrophys.J.}
  {\bfseries 699} (2009)  539--563},
\href{http://arxiv.org/abs/0905.0695}{{\ttfamily arXiv:0905.0695
  [astro-ph.CO]}}.

\bibitem{2011ApJ...730..119R}
A.~G. Riess, L.~Macri, S.~Casertano, H.~Lampeitl, H.~C. Ferguson, A.~V.
  Filippenko, S.~W. Jha, W.~Li, and R.~Chornock, ``{A 3\% Solution:
  Determination of the Hubble Constant with the Hubble Space Telescope and Wide
  Field Camera 3},'' {\em APJ} {\bfseries 730} (2011)  ,
  \href{http://arxiv.org/abs/{1103.2976}}{{\ttfamily arXiv:{1103.2976}
  [astro-ph.CO]}}.

\bibitem{2010JCAP...07..022H}
J.~Hamann, S.~Hannestad, J.~Lesgourgues, C.~Rampf, and Y.~Y.~Y. Wong,
  ``{Cosmological parameters from large scale structure - geometric versus
  shape information},''
  \href{http://dx.doi.org/{10.1088/1475-7516/2010/07/022}}{{\em JCAP}
  {\bfseries 7} (2010)  {22}},
  \href{http://arxiv.org/abs/{1003.3999}}{{\ttfamily arXiv:{1003.3999}
  [astro-ph.CO]}}.

\bibitem{Padmanabhan:2012hf}
N.~Padmanabhan, X.~Xu, D.~J. Eisenstein, R.~Scalzo, A.~J. Cuesta, {\em et al.},
  ``{A 2 per cent distance to $z$=0.35 by reconstructing baryon acoustic
  oscillations - I. Methods and application to the Sloan Digital Sky Survey},''
  \href{http://dx.doi.org/10.1111/j.1365-2966.2012.21888.x}{{\em
  Mon.Not.Roy.Astron.Soc.} {\bfseries 427} (2012) no.~3, 2132--2145},
\href{http://arxiv.org/abs/1202.0090}{{\ttfamily arXiv:1202.0090
  [astro-ph.CO]}}.

\bibitem{Anderson:2012sa}
L.~Anderson, E.~Aubourg, S.~Bailey, D.~Bizyaev, M.~Blanton, {\em et al.},
  ``{The clustering of galaxies in the SDSS-III Baryon Oscillation
  Spectroscopic Survey: Baryon Acoustic Oscillations in the Data Release 9
  Spectroscopic Galaxy Sample},''
  \href{http://dx.doi.org/10.1111/j.1365-2966.2012.22066.x}{{\em
  Mon.Not.Roy.Astron.Soc.} {\bfseries 427} (2013) no.~4, 3435--3467},
\href{http://arxiv.org/abs/1203.6594}{{\ttfamily arXiv:1203.6594
  [astro-ph.CO]}}.

\bibitem{Beutler:2011hx}
F.~Beutler, C.~Blake, M.~Colless, D.~H. Jones, L.~Staveley-Smith, {\em et al.},
  ``{The 6dF Galaxy Survey: Baryon Acoustic Oscillations and the Local Hubble
  Constant},'' \href{http://dx.doi.org/10.1111/j.1365-2966.2011.19250.x}{{\em
  Mon.Not.Roy.Astron.Soc.} {\bfseries 416} (2011)  3017--3032},
\href{http://arxiv.org/abs/1106.3366}{{\ttfamily arXiv:1106.3366
  [astro-ph.CO]}}.

\bibitem{2010ApJ...716..712A}
R.~{Amanullah}, C.~{Lidman}, D.~{Rubin}, G.~{Aldering}, P.~{Astier},
  K.~{Barbary}, M.~S. {Burns}, A.~{Conley}, K.~S. {Dawson}, S.~E. {Deustua},
  M.~{Doi}, S.~{Fabbro}, L.~{Faccioli}, H.~K. {Fakhouri}, G.~{Folatelli}, A.~S.
  {Fruchter}, H.~{Furusawa}, G.~{Garavini}, G.~{Goldhaber}, A.~{Goobar}, D.~E.
  {Groom}, I.~{Hook}, D.~A. {Howell}, N.~{Kashikawa}, A.~G. {Kim}, R.~A.
  {Knop}, M.~{Kowalski}, E.~{Linder}, J.~{Meyers}, T.~{Morokuma}, S.~{Nobili},
  J.~{Nordin}, P.~E. {Nugent}, L.~{{\"O}stman}, R.~{Pain}, N.~{Panagia},
  S.~{Perlmutter}, J.~{Raux}, P.~{Ruiz-Lapuente}, A.~L. {Spadafora},
  M.~{Strovink}, N.~{Suzuki}, L.~{Wang}, W.~M. {Wood-Vasey}, N.~{Yasuda}, and
  T.~{Supernova Cosmology Project}, ``{Spectra and Hubble Space Telescope Light
  Curves of Six Type Ia Supernovae at 0.511 $<$ z $<$ 1.12 and the Union2
  Compilation},'' \href{http://dx.doi.org/{10.1088/0004-637X/716/1/712}}{{\em
  {ApJ}} {\bfseries 716} (2010)  },
  \href{http://arxiv.org/abs/{1004.1711}}{{\ttfamily arXiv:{1004.1711}
  [astro-ph.CO]}}.

\bibitem{Hamann:2007pi}
J.~Hamann, S.~Hannestad, G.~Raffelt, and Y.~Y. Wong, ``{Observational bounds on
  the cosmic radiation density},''
  \href{http://dx.doi.org/10.1088/1475-7516/2007/08/021}{{\em JCAP} {\bfseries
  0708} (2007)  021},
\href{http://arxiv.org/abs/0705.0440}{{\ttfamily arXiv:0705.0440 [astro-ph]}}.

\bibitem{citeulike:5630011}
M.~J.~D. Powell, ``{The BOBYQA algorithm for bound constrained optimization
  without derivatives},'' {\em DAMTP 2009/NA06}  .

\bibitem{Refregier:2010ss}
A.~Refregier, A.~Amara, T.~Kitching, A.~Rassat, R.~Scaramella, {\em et al.},
  ``{Euclid Imaging Consortium Science Book},''
\href{http://arxiv.org/abs/1001.0061}{{\ttfamily arXiv:1001.0061
  [astro-ph.IM]}}.

\bibitem{Corasaniti:2010ze}
P.~S. Corasaniti, ``{The innocuousness of adiabatic instabilities in coupled
  scalar field-dark matter models},''
  \href{http://dx.doi.org/10.1063/1.3462719}{{\em AIP Conf.Proc.} {\bfseries
  1241} (2010)  797--803},
\href{http://arxiv.org/abs/1001.2687}{{\ttfamily arXiv:1001.2687
  [astro-ph.CO]}}.

\bibitem{LaVacca:2009yp}
G.~La~Vacca, J.~Kristiansen, L.~Colombo, R.~Mainini, and S.~Bonometto, ``{Do
  WMAP data favor neutrino mass and a coupling between Cold Dark Matter and
  Dark Energy?},'' \href{http://dx.doi.org/10.1088/1475-7516/2009/04/007}{{\em
  JCAP} {\bfseries 0904} (2009)  007},
\href{http://arxiv.org/abs/0902.2711}{{\ttfamily arXiv:0902.2711
  [astro-ph.CO]}}.

\bibitem{Pettorino:2013oxa}
V.~Pettorino, ``{Testing modified gravity with Planck: the case of coupled dark
  energy},'' \href{http://dx.doi.org/10.1103/PhysRevD.88.063519}{{\em
  Phys.Rev.} {\bfseries D88} (2013) no.~6, 063519},
\href{http://arxiv.org/abs/1305.7457}{{\ttfamily arXiv:1305.7457
  [astro-ph.CO]}}.

\bibitem{Hamann:2007sb}
J.~Hamann, J.~Lesgourgues, and G.~Mangano, ``{Using BBN in cosmological
  parameter extraction from CMB: A Forecast for PLANCK},''
  \href{http://dx.doi.org/10.1088/1475-7516/2008/03/004}{{\em JCAP} {\bfseries
  0803} (2008)  004},
\href{http://arxiv.org/abs/0712.2826}{{\ttfamily arXiv:0712.2826 [astro-ph]}}.

\bibitem{Calabrese:2011hg}
E.~Calabrese, D.~Huterer, E.~V. Linder, A.~Melchiorri, and L.~Pagano, ``{Limits
  on Dark Radiation, Early Dark Energy, and Relativistic Degrees of Freedom},''
  \href{http://dx.doi.org/10.1103/PhysRevD.83.123504}{{\em Phys.Rev.}
  {\bfseries D83} (2011)  123504},
\href{http://arxiv.org/abs/1103.4132}{{\ttfamily arXiv:1103.4132
  [astro-ph.CO]}}.

\bibitem{Albrecht:2006um}
A.~Albrecht, G.~Bernstein, R.~Cahn, W.~L. Freedman, J.~Hewitt, {\em et al.},
  ``{Report of the Dark Energy Task Force},''
\href{http://arxiv.org/abs/astro-ph/0609591}{{\ttfamily arXiv:astro-ph/0609591
  [astro-ph]}}.

\bibitem{Hamann:2012fe}
J.~Hamann, S.~Hannestad, and Y.~Y. Wong, ``{Measuring neutrino masses with a
  future galaxy survey},''
  \href{http://dx.doi.org/10.1088/1475-7516/2012/11/052}{{\em JCAP} {\bfseries
  1211} (2012)  052},
\href{http://arxiv.org/abs/1209.1043}{{\ttfamily arXiv:1209.1043
  [astro-ph.CO]}}.

\bibitem{Basse:2013zua}
T.~Basse, O.~E. Bjaelde, J.~Hamann, S.~Hannestad, and Y.~Y. Wong, ``{Dark
  energy properties from large future galaxy surveys},''
  \href{http://dx.doi.org/10.1088/1475-7516/2014/05/021}{{\em JCAP} {\bfseries
  1405} (2014)  021},
\href{http://arxiv.org/abs/1304.2321}{{\ttfamily arXiv:1304.2321
  [astro-ph.CO]}}.

\bibitem{Basse:2014qqa}
T.~Basse, J.~Hamann, S.~Hannestad, and Y.~Y.~Y. Wong, ``{Getting leverage on
  inflation with a large photometric redshift survey},''
\href{http://arxiv.org/abs/1409.3469}{{\ttfamily arXiv:1409.3469
  [astro-ph.CO]}}.

\end{thebibliography}\endgroup

\end{document}